\documentclass[twocolumn,english,showpacs,preprintnumbers,amsmath,amssymb,floatfix,nofootinbib]{revtex4-1}
\usepackage[T1]{fontenc}
\usepackage[latin9]{inputenc}
\usepackage{color}
\usepackage{array}
\usepackage{amstext}
\usepackage{graphicx}
\usepackage{esint}
\usepackage{cleveref}
\usepackage{rotating}
\usepackage{slashed}
\usepackage[normalem]{ulem}
\usepackage{siunitx}
\usepackage[font=small,labelfont=bf]{caption}

\usepackage{ulem}

\newcommand{\pom} {I\!\!P}

\newcommand{\xpom}{x_{\xpom}}

\makeatletter

\providecommand{\tabularnewline}{\\}

\@ifundefined{textcolor}{}
{%
 \definecolor{BLACK}{gray}{0}
 \definecolor{WHITE}{gray}{1}
 \definecolor{RED}{rgb}{1,0,0}
 \definecolor{GREEN}{rgb}{0,1,0}
 \definecolor{BLUE}{rgb}{0,0,1}
 \definecolor{CYAN}{cmyk}{1,0,0,0}
 \definecolor{MAGENTA}{cmyk}{0,1,0,0}
 \definecolor{YELLOW}{cmyk}{0,0,1,0}
 }

\@ifundefined{definecolor}
 {\usepackage{color}}{}
\@ifundefined{definecolor}
 {\usepackage{color}}{}
\makeatother
\usepackage{babel}

\def\Re{{\cal R \mskip-4mu \lower.1ex \hbox{\it e}\,}}
\def\Im{{\cal I \mskip-5mu \lower.1ex \hbox{\it m}\,}}

\def\tev{\,{\ifmmode\mathrm {TeV}\else TeV\fi}}
\def\gev{\,{\ifmmode\mathrm {GeV}\else GeV\fi}}
\def\mev{\,{\ifmmode\mathrm {MeV}\else MeV\fi}}
\def\to{\rightarrow}

\begin{document}

%%%%%%%%%%%%%%%%%%%%%%%%%%%%%%

\title { Phenomenology of diffractive DIS in the framework of fracture functions and determination of diffractive parton distribution functions  }

\author{Hamzeh Khanpour}
\email{Hamzeh.Khanpour@mail.ipm.ir}

\affiliation {
Department of Physics, University of Science and Technology of Mazandaran, P.O.Box 48518-78195, Behshahr, Iran    \\ 
School of Particles and Accelerators, Institute for Research in Fundamental Sciences (IPM), P.O.Box 19395-5531, Tehran, Iran 
}

\date{\today}

%
%%%%%%%%%%%%%%%%%%%%%%%%%%%%%%%%%%%%%%%%%%%%%%%%%%%%%%%%%%%%%%%%%%%%%%%%%%%%%%%%%%%%%%%%%%%%%%%%%%%%%%%
\begin{abstract}\label{abstract}

The goal of this study is to determine a set of diffractive parton distribution function (diffractive PDFs) from a QCD analysis of all available and up-to-date diffractive deep inelastic scattering (diffractive DIS) data sets from HERA $ep$ collider, including the most recent H1 and ZUES combined inclusive diffractive cross section
measurements. This extraction of diffractive PDFs, referred to as {\tt HK19-DPDF}, is performed at next-to-leading (NLO) and next-to-next-to-leading (NNLO) in perturbative QCD. This new determination of diffractive PDFs is based on the fracture functions methodology, a QCD framework designed to provide a statistically sound representation of diffractive DIS processes. Heavy quark contributions to the diffractive DIS are considered within the framework of the FONLL general mass variable flavor number scheme (GM-VFNS) and the ``Hessian approach'' is used to determine the uncertainties of diffractive PDFs.
We discuss the novel aspects of the approach used in the present analysis, namely an optimized and flexible parametrization of the diffractive PDFs as well as a strategy based on the fully factorization theorem for diffractive hard processes.
We then present the diffractive PDFs, and discuss the fit quality and the stability upon variations of the kinematic cuts and the fitted data sets. We find that the systematic
inclusion of higher-order QCD corrections could improves the description of the data. We compare the extracted sets of diffractive PDFs based on the fracture functions approach to other recent sets of diffractive PDFs, finding in general very good agreements.

\end{abstract}
%

%\pacs{12.39.-x, 14.65.Bt, 12.38.-t, 12.38.Bx}

\maketitle
\tableofcontents{}

%
%%%%%%%%%%%%%%%%%%%%%%%%%%%%%%%%%%%%%%%%%%    Introduction    %%%%%%%%%%%%%%%%%%%%%%%%%%%%%%%%%%%%%%%%%%%%%%%%%
\section{Introduction}\label{sec:intro}

In the past decades, there has been an increasing interest in attempting to understand the structure of hadron~\cite{Gao:2017yyd,Newman:2013ada}.
The high energy processes of interest which contain information on the hadron structure in Quantum Chromodynamics (QCD) mostly include the hadron production in in lepton-nucleon ($\ell p$) deep-inelastic scattering (DIS) and in proton-proton ($pp$) collisions. Information from hadron collisions, especially at the large hadron collider (LHC) at CERN, is particularly useful in order to achieve a precise understanding of non-perturbative QCD dynamics.
Since its start of data taking, the H1 and ZEUS experiments at HERA-I and HERA-II have provided an impressive wealth of information on the
quark and gluon structure of the proton, and hence, a considerable amount of literature has been published over the past decade. Indeed, modern global analyses of parton distribution functions (PDFs)~\cite{H1:2018flt,Ball:2014uwa,Accardi:2016qay,Harland-Lang:2014zoa} as well as diffractive PDFs include a wide range of HERA measurements of electron-proton process. Recent developments in the field of PDFs of the proton have led to a renewed interest in diffractive DIS process to extract the non-perturbative diffractive PDFs from a global analysis. Significant progresses in understanding diffraction processes have been made at the HERA collider, where typically a 27.5 GeV electron (or positron) collides with a 820 or 920 GeV proton. Recent analyses of diffractive DIS data collected by H1 and ZEUS collaborations at HERA have confirmed substantial contributions of perturbative QCD-based effects in diffractive DIS cross sections. In general, in $ep$ collisions at HERA, the hard diffraction contributes a fraction of order 8-10\% to the total DIS cross sections.

Future high precision and high energy DIS experiments are expected to reach a wider kinematic range of momentum fraction $x$ and photon virtuality Q$^2$ which could not be explored previously by $ep$ HERA. Among them are the Large Hadron-electron Collider (LHeC)~\cite{Klein:2018rhq} and Future Circular Collider in electron-hadron mode (FCC-eh)~\cite{Mangano:2018mur,Benedikt:2018csr}. LHeC would utilize the 7 TeV proton beam from the LHC and collides it with a 60 GeV electron beam, and would extend the available kinematic range in $x$ and photon virtuality Q$^2$ by a factor of order 20 and 100, respectively. Beyond the LHeC, the next generation $ep$ collider FCC-eh, utilizing the 50 TeV proton beam from the FCC which would probe DIS processes at center-of-mass energy of 3.5 TeV, much higher than the HERA collider, leading to a better understanding of the proton structure with extremely high precision. Recently, an investigated of the potential of theses high energy and high luminosity machines for the measurement of diffractive DIS cross sections and to constrain the diffractive PDFs has been performed, see the analysis of Ref.~\cite{Armesto:2019gxy} for details.

On the phenomenological side, a similar strategy to determine the PDFs can also be adapted to the case of diffractive DIS, considering the collinear factorization~\cite{Collins:1997sr} and the validity of proton vertex factorization~\cite{Ingelman:1984ns}. Diffractive PDFs can be extracted from QCD analyses of diffractive DIS data sets. Similarly to the PDFs, the diffractive PDFs are expected to obey the by the (Dokshitzer?Gribov?Lipatov?Altarelli?Parisi) DGLAP equations. In this picture of proton vertex factorization, the diffractive DIS processes are described by the exchange of colourless object such a pomeron. Recent progress in the determination of diffractive PDFs widely used the diffractive DIS data sets from H1 and ZEUS experiments at HERA (see, for example, Refs.~\cite{Armesto:2019gxy,Helenius:2019gbd,Rasmussen:2018dgo} for recent reviews). In the last few years, at least three groups have reported sets of diffractive PDFs with uncertainties using the mentioned data sets: {\tt H1-2006-DPDF}~\cite{Aktas:2006hy}, {\tt ZEUS-2010-DPDF}~\cite{Chekanov:2009aa}, and the most recent analysis by {\tt GKG18-DPDF}~\cite{Goharipour:2018yov}. All these of diffractive PDFs determinations were performed at next-to-leading order (NLO)
accuracy in perturbative QCD. The primary focus of these QCD analyses were put on quantifying the effects of the inclusion of new measurements of diffractive DIS at HERA as well as the diffractive dijet production. The analysis by {\tt GKG18-DPDF} introduced some improvements over previous determinations. Specifically, in order to achieve a more reliable estimate of the diffractive PDFs uncertainties, the {\tt xFitter} package have been employed. In addition, {\tt GKG18-DPDF} used for the first time the most recent H1/ZEUS combined diffractive DIS cross section measurements. Up to now, predictions for diffractive DIS, and in particular for diffractive dijet production, were performed only at next-to-leading order (NLO) accuracy. Recently, predictions for the diffractive dijet production, also is provided at next-to-next-to-leading order (NNLO) in Ref.~\cite{Britzger:2018zvv}. Although much theoretical work remains to be done for this process, it is already clear that the accumulation of precise data for diffractive DIS processes will greatly deepen our understanding of perturbative QCD (pQCD).

In the present study we construct for the first time a set of NLO and NNLO diffractive PDFs using all available and up-to-date diffractive DIS cross section measurements~\cite{Aaron:2012ad,Aaron:2012zz} including the most recent H1 and ZEUS combined measurement for the inclusive diffractive DIS cross sections~\cite{Aaron:2012hua}. We do so by using a methodology which has been suggested in Refs.~\cite{Trentadue:1993ka,deFlorian:1998rj}, and recently used to study leading neutron production~\cite{Ceccopieri:2014rpa,Shoeibi:2017lrl,Shoeibi:2017zha} as well as inclusive diffractive DIS~\cite{Ceccopieri:2016rga}: the so called ``fracture function approach''.

Similar to the case of ordinary structure functions in totally inclusive DIS, QCD does not predict the shape of the fracture functions unless it is known at
a certain initial scale, Q$_0^2$. This universal and non-perturbative information has to be determined from a QCD analysis of experimental data sets, and hence, can be parametrized finding inspiration in non-perturbative models. More recently, the fracture functions approach has been successfully applied to describe leading neutron and leading proton productions at H1 and ZEUS experiment, considering a model as non-perturbative input~\cite{Ceccopieri:2014rpa,Shoeibi:2017lrl,Shoeibi:2017zha}.

The main aim of this analysis is to provide a conceptual theoretical framework based on fracture function approach to describe the diffractive DIS processes. 
As a result of the following analysis, we present a parametrization for the fracture function that characterizes the underlying diffractive DIS process at
an initial scale, as extracted from H1 and ZEUS data sets. The obtained parametrization is used to compute other observables measured by H1, not included in our QCD fit, finding also an outstanding agreement with the data. In this analysis we emphasis that an approach in the framework of fracture functions phenomenologically allows a accurate and reliable description of diffractive DIS cross sections. Our results also verify that the scale dependence of the diffractive DIS data agrees well with the one predicted by the use of fracture function formalism. Therefore, this paper provides an important opportunity to advance the understanding of diffractive DIS events measured by H1 and ZEUS collaboration at HERA.

This paper has been organized in the following way: Sec.~\ref{sec:framework} begins by laying out the theoretical framework and assumptions of this research. In
particular, we discuss the methodology used for this study including the diffractive structure functions, diffractive PDFs, hard-scattering factorization and heavy quark contributions. Then in Sec.~\ref{sec:Input} we present our input for the diffractive PDFs at a given initial scale $Q^2 = Q_0^2$. In Sec.~\ref{sec:data} we concentrate on the description of the H1 data sets which have been added in our QCD analysis, together with a discussion on the inclusion of H1 and ZEUS combined data sets. 
Sec.~\ref{sec:minimizations} is concerned with the methodology used in this study to determine the diffractive PDFs uncertainty. The results of the global analysis can be found in Sec.~\ref{sec:FitResults}. This section starts with a presentation of our NLO and NNLO diffractive PDFs and their uncertainties, together with the values of the input parameters. Detailed discussions of the main results and comparisons with the analyzed data sets are also presented as well. Finally, the conclusion in Sec.~\ref{sec:Discussion} gives a brief summary and critique of the findings.

%
%%%%%%%%%%%%%%%%%%%%%%%%%%%%%%%%%%%%%%%%%%%%%%%%%%%%%%%%%%%%%%%%%%%%%%
\section{Theoretical framework and assumptions}\label{sec:framework}
%%%%%%%%%%%%%%%%%%%%%%%%%%%%%%%%%%%%%%%%%%%%%%%%%%%%%%%%%%%%%%%%%%%%%%
%

Predictions for the diffractive processes in DIS can be obtained in the framework of perturbative QCD (pQCD). As we already mentioned, hard processes in diffractive DIS can be described by a factorization into parton level subprocesses and diffractive PDFs. In this framework, cross sections for diffractive deep inelastic lepton proton ($\ell p$) scattering can be computed at NLO and NNLO accuracy in pQCD.

%
%%%%%%%%%%%%%%%%%%%%%%%%%%%%%%%%%%%%%%%%%%%%%%%%%%%%%%%%%%%%%%%%%%%%%%
\subsection{ Diffractive structure functions }\label{sec:SF}
%%%%%%%%%%%%%%%%%%%%%%%%%%%%%%%%%%%%%%%%%%%%%%%%%%%%%%%%%%%%%%%%%%%%%%
%

\begin{figure}[htb]
\vspace{0.250cm}
\includegraphics[clip,width=0.48\textwidth]{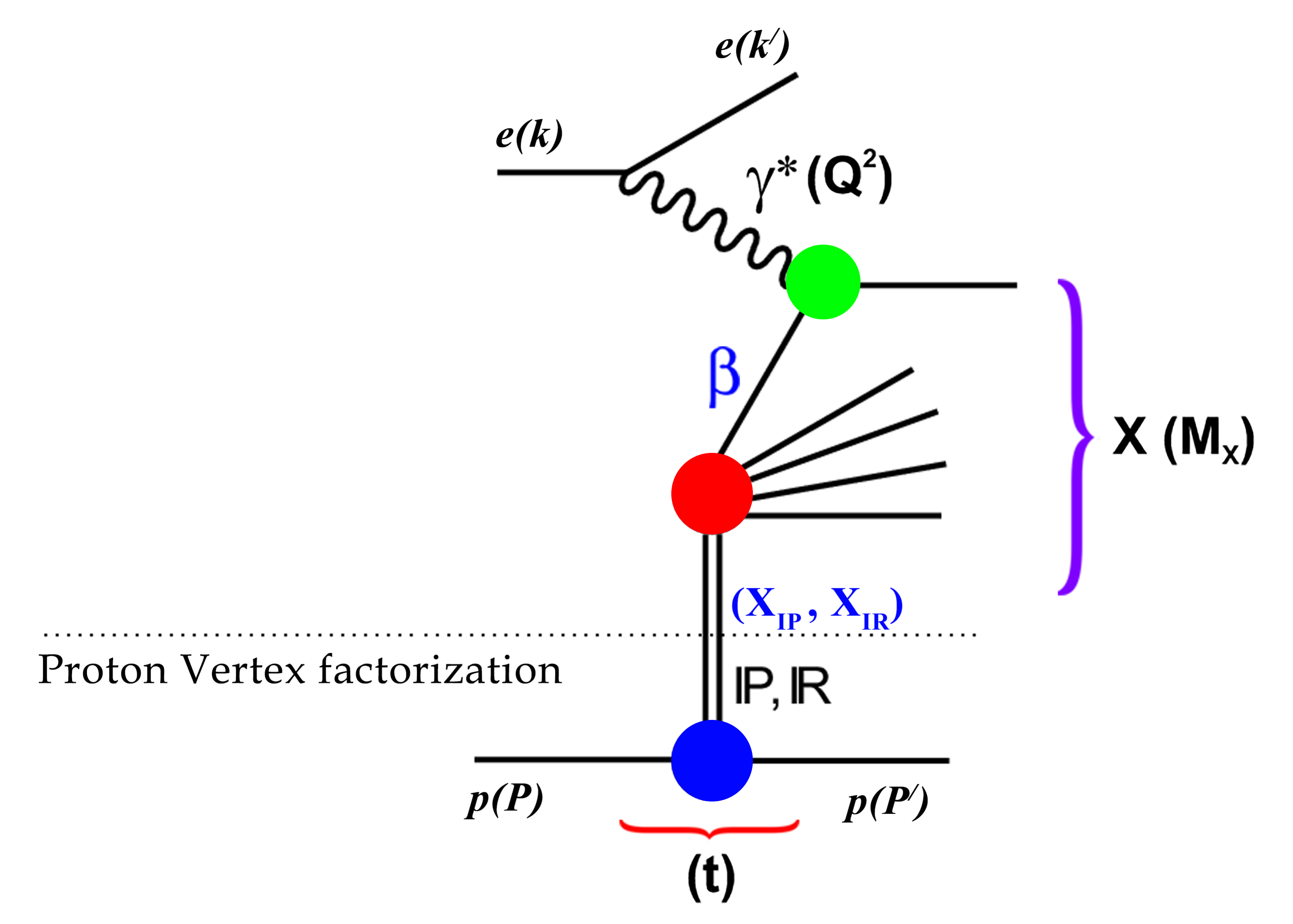}
%\vspace{-6cm}
\begin{center}
\caption{{\small Representative Feynman diagram for the neutral current diffractive DIS process $e p \to e p X$ proceeding via virtual photon exchange in the picture of proton vertex factorization. The kinematic variables are described in the text. } \label{fig:Feynman-Old}}
\end{center}
\end{figure}

\begin{figure}[htb]
\vspace{0.250cm}
\includegraphics[clip,width=0.48\textwidth]{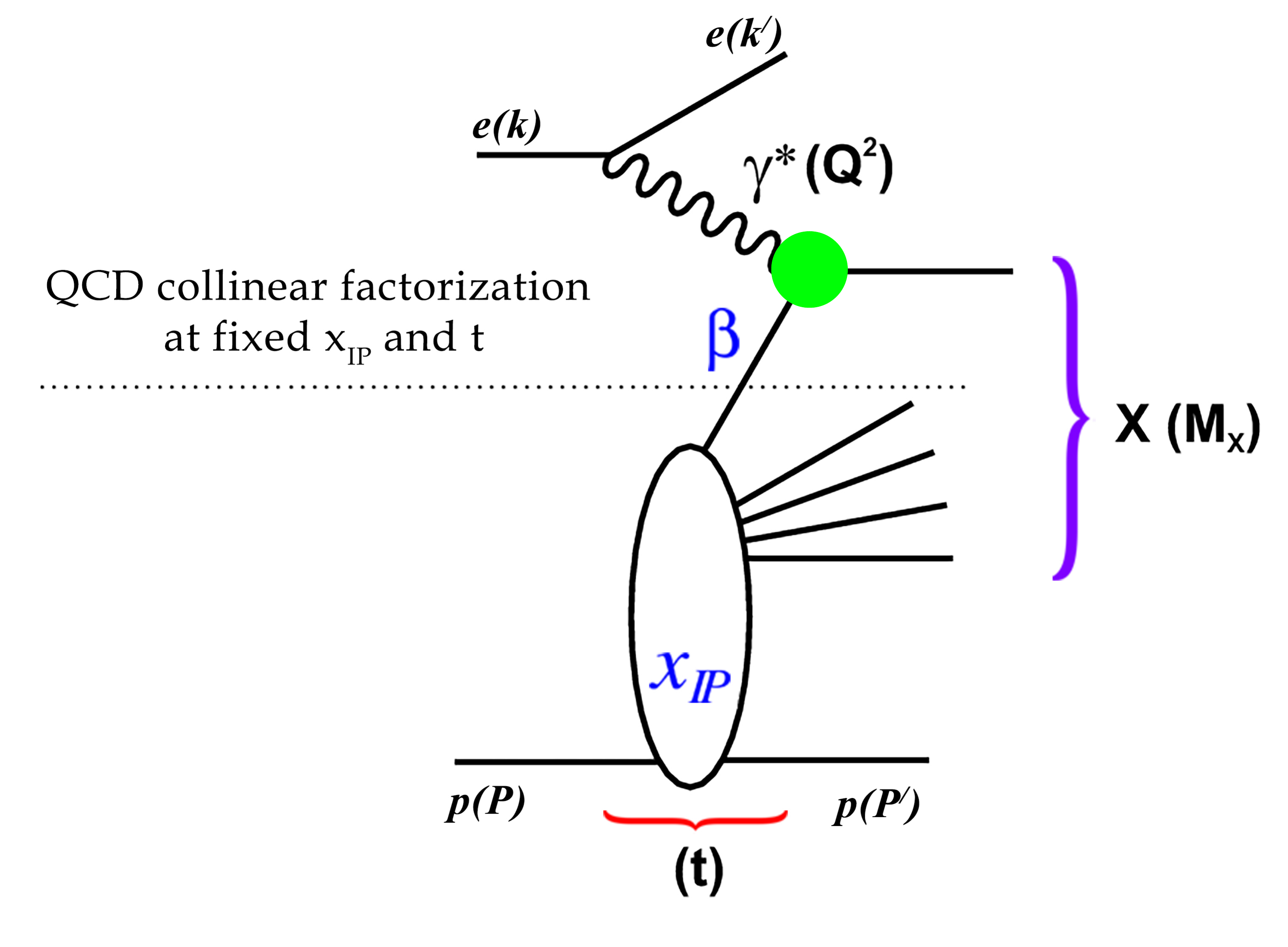}
%\vspace{-6cm}
\begin{center}
\caption{{\small Representative Feynman diagram for the neutral current diffractive DIS process $e p \to e p X$ proceeding via virtual photon exchange in the picture of QCD hard scattering collinear factorization. The kinematic variables are described in the text. } \label{fig:Feynman-New}}
\end{center}
\end{figure}

The kinematical variables to describe the diffractive DIS events can be inferred from the momenta of the incoming lepton $(\ell )$, proton $(p)$ and the outgoing lepton. The leading order Feynman diagram for diffractive DIS at HERA is displayed in Figs.~\ref{fig:Feynman-Old} and ~\ref{fig:Feynman-New} in the picture of proton vertex and collinear and QCD hard scattering
collinear factorizations, respectively. In the diffractive DIS in which belongs to the semi-inclusive lepton proton DIS (SIDIS), 

%------------------------------------------------
%------------------------------------------------

%------------------------------------------------
%------------------------------------------------

%--------------------------------------
\begin{eqnarray}\label{proc}
\ell (k)  +  p(P) \to \ell (k^{\prime})
+  p(P^{\prime})
+  X(p_{X})  \,,
\end{eqnarray}
%--------------------------------------

in which in addition to the outgoing leptons, an additional proton $(p)$ can be detected in the final state of the diffractive DIS processes.
In above equation, $X$ indicates to the unobserved part of the hadronic final state. The kinematics of such events is specified by the following kinematical variables: 

%--------------------------------------
\begin{eqnarray}\label{kinematics}
Q^{2} = -q^{2},
\, \, \,  \,
x_{B} = \frac{Q^2}{2P.q},
\, \, \, \,
 y = \frac{P.q}{P.k} \,.
\end{eqnarray}
%--------------------------------------

The momentum transfered to the target proton  $(p)$  is given by the momentum $q^2 = (k - k^{\prime})^{2}$ of the virtual photon $\gamma^*$.
Hence, the $Q^{2}$ in Eq.~\eqref{kinematics} is the well-known photon virtuality. $x_{B}$ is the usual $x$-Bjorken variable in DIS process which idicate the longitudinal momentum fraction of the proton carried by the struck quarks, and ${y}$ is refereed to the inelasticity of the DIS scattering.

In the lepton-proton ($\ell p$) center-of-mass (CM) system, diffractive DIS processes are interpreted by an outgoing proton ${(p)}$ with a large momentum fractions of the incident protons as well as quite small values of the  invariant transverse momentum ${t}$ measured with respect to the collision axis in the ``target fragmentation'' regions of the incident protons.

In order to the describe such diffractive events, one need to introduce additional kinematics invariants. As illustrated in Fig.~\ref{fig:Feynman-New}, the longitudinal
momentum fractions, $x_{\pom}$ of the colourless exchange with respect to the incoming proton ${(p)}$ momentum, and the invariant momentum transfer ${t}$ at the proton vertex ${t}$, and ${\beta}$ of the struck quarks with respect to the colourless exchange, are also needed to describe the diffractive DIS processes. These two new variables are given by,

%--------------------------------------
\begin{eqnarray}
x_{\pom} = \frac{q.(P - P^\prime)} {P.q}, \, \, \,  \, \beta = \frac{Q^2} {Q^2 + M_X^2} \,.
\end{eqnarray}
%--------------------------------------

$M_{X}$ is the invariant mass of the hadronic final state ${X}$. The scaled fractional momentum variable ${\beta}$ which is also defined by ${\beta} = \frac{x_B}{x_{\pom}}$ interpreted as the fractional momentum of interacting partons in the protons with respect to the pomeron ${\pom}$ fractional momentum ${x_{\pom}}$. 

Now we turn to inclusive diffractive DIS cross sections. The neutral current (NC) measurements in diffractive DIS are often presented in terms of the reduced lepton-proton ($\ell p$) cross section, ``neutral current diffractive reduced cross section'' $\sigma_r^{D(4)}$. The $t$-unintegrated reduced cross section depends on the diffractive transverse and longitudinal structure functions $F_2^{D(4)}$ and $F_L^{D(4)}$, respectively. In the one-photon exchange approximation, it is given by:

%--------------------------------------
\begin{eqnarray}\label{sigma-r}
\sigma_r^{D(4)} (\beta, Q^2; x_{\pom}, t) = && F_2^{D(4)}  (\beta, Q^2; x_{\pom}, t)  - \nonumber \\
&& \frac{y^2}{1+(1-y)^2} F_L^{D(4)}  (\beta, Q^2; x_{\pom}, t) \,. \nonumber \\
\end{eqnarray}
%--------------------------------------

In the measurements of inclusive diffractive DIS at HERA, the data were presented in terms of $\sigma_r^{D(4)} (\beta, Q^2; x_{\pom})$.
In this analysis, we analyze all available and up-to-date diffractive DIS data sets~~\cite{Aaron:2012zz,Aaron:2012ad} including the most recent data for combined H1 and ZEUS diffractive DIS cross sections measurements~\cite{Aaron:2012hua}.

%
%%%%%%%%%%%%%%%%%%%%%%%%%%%%%%%%%%%%%%%%%%%%%%%%%%%%%%%%%%%%%%%%%%%%%%
\subsection{ Diffractive parton distributions and hard-scattering factorization }\label{sec:factorization}
%%%%%%%%%%%%%%%%%%%%%%%%%%%%%%%%%%%%%%%%%%%%%%%%%%%%%%%%%%%%%%%%%%%%%%
%

According to the factorization theorem for diffractive DIS in pQCD~\cite{Collins:1997sr,Collins:2001ga,Grazzini:1997ih}, if the process is sufficiently hard, the calculation
can be subdivided into two components: the hard partonic cross sections which are calculable within the pQCD in powers of strong coupling constant ${\alpha_{s}}$, which need to be convoluted with soft diffractive PDFs, ${\cal F}_{i/p}^{D}  (\beta, \mu^{2}_{F}; x_{\pom}, t)$ that specify the contributing partons inside the incoming hadrons. Likewise the PDFs, diffractive PDFs are also universal for all diffractive DIS events  with the hardness of the process being ensured by the virtuality $Q^{2}$ of the exchanged photons. The structure functions (SF) appearing in Eq.~\eqref{sigma-r} are given by

%--------------------------------------
\begin{eqnarray}\label{eq:factorization}
F_{k}^{D(4)} (\beta, Q^{2}; x_{\pom}, t)
= &&  \sum_{i} \int_{\beta}^{1}
\frac{d \zeta}{\zeta} {\cal F}_{i/p}^{D}
(\beta, \mu^{2}_{F}; x_{\pom}, t)  \nonumber \\
&& C_{ki} \big (\frac{\beta} {\zeta},
\frac{Q^{2}}{\mu^{2}_{F}}, \alpha_s(\mu^{2}_{R}) \big)
+ {\cal O} (\frac{1} {Q^{2}}) \,.  \nonumber \\
\end{eqnarray}
%--------------------------------------

The index $i$ in Eq.~\eqref{eq:factorization} runs on the flavors of the interacting partons in the proton~\cite{Ceccopieri:2016rga}. The quark and gluon hard scattering coefficient functions, $C_{q}$ and $C_{g}$, which appeared in above equation are perturbatively
calculable as a power expansion in the strong coupling and are the same as in fully inclusive DIS process~\cite{Vermaseren:2005qc}.  As one can find in literature, for example~\cite{Goharipour:2018yov,Martin:2004xw,Martin:2005hd,Monfared:2011xf,Chekanov:2009aa,Aktas:2006hy,Aaron:2012zz,Aaron:2012ad,Aaron:2012hua} as well as our discussion in previous section, in order to describe the diffractive DIS events, two more variable in addition to the usual DIS kinemtical variables are also needed which are the longitudinal momentum fraction $x_{\pom}$ and the invariant momentum transfer ${t}$.
In the general factorization theorem~\cite{Collins:1997sr,Collins:2001ga,Grazzini:1997ih} for the diffractive DIS events in the form of above equation holds at fixed values of $x_{\pom}$ and invariant momentum transfer ${t}$~\cite{Ceccopieri:2016rga}.
Hence, the parton content of the colour singlet exchange described by diffractive PDFs, ${\cal F}_{i/p}^{D}$, is uniquely controlled by the kinematics of the outgoing protons in which carry the fractional momentum of $1 - x_{\pom}$.

The diffractive PDFs, ${\cal F}_{i/p}^{D}(\beta, \mu^2_{F}; x_{\pom}, t)$, appearing in Eq.~\eqref{eq:factorization} are known as ``proton-to-proton fracture functions''~\cite{Trentadue:1993ka} in the very forward kinematical region.
It can be interpreted as the number density of interacting partons at a given scale of $\mu^{2}$ with fractional momentum of ${\beta}$ conditional to the detection of a final state proton $(p)$ with fractional momentum $1-x_{\pom}$ and invariant momentum transfer ${t}$~\cite{Ceccopieri:2016rga}. 
As in the inclusive case, the $Q^{2}$ evolution of the diffractive DIS events can be predicted in pQCD. Since the scale dependence of the cross section in leading particle production in DIS processes can be calculated within pQCD~\cite{Trentadue:1993ka,Ceccopieri:2007th}, therefore the diffractive PDFs also obey the standard DGLAP evolution equations~\cite{Ceccopieri:2014rpa,Camici:1998bg,Ceccopieri:2007th,Shoeibi:2017lrl,Shoeibi:2017zha}. The ${t}$-unintegrated diffractive PDFs presented in Eq.~\eqref{eq:factorization} also obey the standard DGLAP evolution equations~\cite{Ceccopieri:2014rpa,Ceccopieri:2016rga}. 

By integrating Eq.~\eqref{eq:factorization} with respect to the momentum transfer ${t}$ over up to values of order Q$^{2}$, the diffractive PDFs obey an inhomogeneous DGLAP-type evolution equations~\cite{Trentadue:1993ka}.
We should note here that most of diffractive DIS events occur for the small values of ${t}$. In the case for the H1/ZEUS combined data, the squared four-momentum transfer at the proton vertex, ${t}$, is integrated in the restricted range of $0.09~GeV^2 < |t| < 0.55 ~ GeV^2$ . In the case where the diffractive PDFs presented in Eq.~\eqref{eq:factorization} are integrated over ${t}$ in a limited range, they can obey the standard DGLAP evolution equation~\cite{Ceccopieri:2007th,Ceccopieri:2016rga}.  Hence, we considered such approximation in our analysis and therefore one can use the standard DGLAP evolution when ${\cal F}_{i/p}^{D}  (\beta, \mu^2_F; x_{\pom}, t)$ are integrated over ${t}$ in a limited range~\cite{Ceccopieri:2007th,Ceccopieri:2016rga}

%--------------------------------------
\begin{eqnarray}\label{eq:t-integrated}
{\cal F}_{i/p}^D  (\beta, \mu^2_F; x_{\pom}) =  \int_{t_{\rm min}}^{t_{\rm max}} dt \, {\cal F}_{i/p}^D  (\beta, \mu^2_F; x_{\pom}, t)\,,  \,  \, \, t_{\rm max} \ll Q^2\,. \nonumber \\
\end{eqnarray}
%--------------------------------------

Hence, like for the case of PDFs, the evolution equations of diffractive PDFs in diffractive DIS are easily obtained by the DGLAP evolution equations~\cite{Trentadue:1993ka,Ceccopieri:2007th} as

%--------------------------------
\begin{eqnarray}\label{eq:DGLAP}
Q^2 \frac{\partial {\cal F}^D_{\Sigma/p} (\beta, Q^2; x_{\pom})}{\partial Q^2}  &&  =  \frac{\alpha_s(Q^2)}{2 \pi}  \nonumber \\ &&  \int_{\beta}^{1} \frac{du}{u} P_{\Sigma}^j(u) \, {\cal F}^D_{\Sigma/p} (\frac{\beta }{u}, Q^2; x_{\pom})\,, \nonumber  \\
Q^2 \frac{\partial {\cal F}^B_{g/p} (\beta, Q^2; x_{\pom})}{\partial Q^2}    &&  =  \frac{\alpha_s(Q^2)}{2 \pi} \nonumber \\ && \int_{\beta}^{1} \frac{du}{u} P_{g}^j(u) \, {\cal F}^D_{g/p} (\frac{\beta }{u}, Q^2; x_{\pom})\,,
\end{eqnarray}
%--------------------------------

where ${\cal F}^D_{\Sigma/p} (\beta, Q^2; x_{\pom})$ and ${\cal F}^B_{g/p} (\beta, Q^2; x_{\pom})$ correspond to the singlet and gluon diffractive distributions, respectively.
These non-perturbative distributions need to be parametrize at an input scale Q$_0^2$ and their evolution to higher scale, $Q^2>Q_0^2$, can be described by using the evolution equation given above.
$P_{\Sigma}$ and $P_{g}$ in Eq.~\eqref{eq:DGLAP} are the common NLO and NNLO contributions to the splitting functions governing the evolution of unpolarized singlet $(S)$ and non-singlet $(NS)$ combinations of quark densities in perturbative QCD.
Splitting functions are perturbatively calculable as a power expansion in the strong coupling constant $\alpha_{s}$. The splitting functions $P_{\Sigma}$ and $P_{g}$ in Eq.~\eqref{eq:DGLAP} are the same as in fully inclusive DIS~\cite{Vogt:2004mw,Moch:2004pa}.

We have now, in principle, all the essential ingredients to write down the diffractive DIS cross sections of Eq.~\eqref{sigma-r} in terms of diffractive structure functions.
In the next sections, we give a detailed account of the global analysis of diffractive PDFs performed in this study. We first discuss the heavy flavour contributions, the details parameterization of diffractive PDFs and then we will present data selection and the determination of the best fit, which we compare to the fitted data. We then focus on the studies of uncertainties using the standard ``Hessian'' error matrix approach.

%=====================================================================================
\subsection{ Heavy flavour contributions to the diffractive DIS structure function }\label{sec:heavy-flavour}
%=====================================================================================

Is is shown in literature that the correct treatments of heavy flavors have an important impact on the PDFs extracted from the global analysis due to the data available for
heavy flavor structure functions, $F^h_L (x, Q^2, m_h^2)$ with $h = c, b$ and $k= 2, L$, as well as the heavy flavor contributions to the total inclusive structure functions at small values of momentum fraction $x$~\cite{Ball:2014uwa,Martin:2009iq,Alekhin:2017kpj,Bertone:2017djs}.
There are several scheme available in literature where heavy quark production can be readily described. Among them are zero-mass variable flavor scheme (ZM-VFNS), fixed-flavor number scheme (FFNS) and general-mass variable flavour number scheme (GM-VFNS). 

The resulting calculations for the heavy quark coefficient functions are more accurate in two different and complementary regimes. The FFNS is more accurate for values of Q$^2$ comparable to the mass of the heavy quark involved in the calculation, while the ZM-VFNS is instead more accurate for values of Q$^2$ much larger than the heavy-quark mass, $Q^2 \gg m_h^2$. It has been shown in literature that the GM-VFNS is the most appropriate scheme to extract PDFs from a global QCD analysis~\cite{Thorne:2012az,Ball:2013gsa,Thorne:2014toa}. 
This scheme could extrapolates smoothly from the FFNS at low value of Q$^2$ to the ZM-VFNS at high Q$^2$, and therefore, produces a good description of the effect of heavy quark contributions to the structure functions as well as inclusive cross section over the whole range of Q$^2$. Well known examples of GM-VFNS are the original Aivazis-Collins-Olness-Tung ({\tt ACOT GM-VFNS}) scheme~\cite{Aivazis:1993kh}, the Thorne-Roberts ({\tt TR GM-VFNS}) scheme~\cite{Thorne:1997ga} and the so-called {\tt FONLL GM-VNFS} scheme~\cite{Forte:2010ta}. In our previous diffractive PDFs analysis, {\tt GKG18-DPDF}~\cite{Goharipour:2018yov} we adopted the more recent ``{\tt TR}'' prescription~\cite{Thorne:2006qt}.

In the present analysis, we prefer to use the {\tt FONLL GM-VFNS}~\cite{Forte:2010ta} which provides the charm structure functions by using the exact $x$-space ${\cal O} (\alpha_s^2)$ heavy quark coefficient functions as well as computes neutral- and charged-current DIS observables up to this order. The evolution of diffractive PDFs ${\cal F}^D (\beta, Q^2; x_{\pom})$ is performed within the FONLL GM-VFNS by using the publicly available {\tt APFEL} program~\cite{Bertone:2013vaa}. The QCD parameters are the ones quoted in the analysis by {\tt NNPDF} Collaboration~\cite{Bertone:2018ecm,Bertone:2017tyb}. In particular, we use the heavy flavor masses for charm and bottom as $m_c = 1.51$ GeV and $m_b = 4.92$ GeV, respectively. Also the $Z$-boson mass is chosen to be $M_Z = 91.1876$ GeV and the the strong coupling is evaluated at two loop setting $\alpha_{s, \rm NLO}^{n_F=5} (M_Z) =  0.1185$~\cite{Tanabashi:2018oca}. This selection of $\alpha_{s} (M_Z)$ is consistent with the very recent determination of the strong coupling constant reported by the {\tt NNPDF3.1}~\cite{Ball:2018iqk}, in which for the first time uses the jet production and $t \bar t$ differential distributions at NNLO accuracy.

%
%%%%%%%%%%%%%%%%%%%%%%%%%%%%%%%%%%%%%%%%%%%%%%%%%%%%%%%%%%%%%%%%%%%%%%
\section{ Input distributions }\label{sec:Input}
%%%%%%%%%%%%%%%%%%%%%%%%%%%%%%%%%%%%%%%%%%%%%%%%%%%%%%%%%%%%%%%%%%%%%%
%

In this section, we present the input distributions for the diffractive PDFs at the reference scale of $Q^2 = Q_0^2$. We also glance ahead to mention some of the main feature of this selection.
As is clear from the discussion in Sec.~\ref{sec:factorization}, one improvement in the following analysis is to use parameterizations for the input diffractive PDFs based on the fracture function approach. Following the detailed studies in Refs.~\cite{Shoeibi:2017zha,Shoeibi:2017lrl}, we take for most PDFs a parameterization of the form

%--------------------------------------
\begin{eqnarray}\label{eq:DPDF-Q0-1}
\beta \, {\cal F}_q (\beta, Q_0^2;x_{\pom})  =  {\cal W} (x_{\pom}) ~   \beta \, f_q (\beta, Q_0^2) \,,  \nonumber \\
\beta \, {\cal F}_g (\beta, Q_0^2;x_{\pom})  =  {\cal W} (x_{\pom}) ~   \beta \, f_g (\beta, Q_0^2) \,,
\end{eqnarray}
%--------------------------------------

where $Q_0^2 = 2$ GeV$^2$ is the the input scale, and  ${\cal W} (x_{\pom})$ is the flux factor which we assume that it depends only to the $x_{\pom}$ variable. For the diffractive flux factor, we use the standard functional form which can be written as,

%--------------------------------------
\begin{eqnarray}\label{eq:Flux-Q0}
{\cal W} (x_{\pom}) = x_{\pom}^{w_1} (1 - x_{\pom})^{w_2} (1 + w_3 x_{\pom}^{w_4}) \,. 
\end{eqnarray}
%--------------------------------------

The global fit determines the values of the set of parameters $w_1$, $w_2$, $w_3$ and $w_4$ for diffractive quark and gluon PDFs. 
We consider that this flux factor to be the same for the light sea and gluon, as there is not enough data which can constrain these distributions, while leaving all four parameters $w_i$ in the polynomial free leads to instabilities in the fit. The dependence of the diffractive PDFs on the $x_{\pom}$ depends very much on the parameterization that one chooses for the flux factor. In Eq.~\eqref{eq:DPDF-Q0-1}, $\beta \, f_q (\beta, Q_0^2)$ and $\beta \, f_g (\beta, Q_0^2)$ are the quark and gluon densities at the reference scale $Q_0^2$. The quark and gluon densities depend on the scaled fractional momentum variable $\beta$ and photon virtuality $Q^2$. 

%--------------------------------------
\begin{eqnarray}\label{eq:DPDF-Q0-2}
\beta \, f_q (\beta, Q_0^2) = {\cal N}_q \, \beta^{\alpha_q} (1 - \beta)^{\beta_q} ( 1 + \gamma_q \sqrt{\beta} + \eta_q \beta^2 )\,, \nonumber \\
\beta \, f_g (\beta, Q_0^2) = {\cal N}_g \, \beta^{\alpha_g} (1 - \beta)^{\beta_g} ( 1 + \gamma_g \sqrt{\beta} + \eta_g \beta^2 )\,.  \nonumber \\
\end{eqnarray}
%--------------------------------------

For the quark density $\beta \, f_q (\beta, Q_0^2)$, we consider a standard functional parameterization. Not all the parameters in our inputs for the quark density are free. We will return to this issue that due to the lack of enough diffractive DIS data sets, one needs to fixed some of these variables at their best fit values, and therefore, there are potentially less free parameters in the diffractive PDFs fit rather than what we presented in Eq.~\eqref{eq:DPDF-Q0-2}.
Like to the case of quark density, the poorly determined gluon density $\beta \, f_g (\beta, Q_0^2)$ is also taken to has a simpler input functional form as presented in Eq.~\eqref{eq:DPDF-Q0-2}.
It is worth to mention here that our parametrizations for the input distributions and all free parameters listed there allow the QCD fits a large degree of flexibility. 

Considering the flux factor introduced in Eq.~\eqref{eq:Flux-Q0} as well as the quark and gluon densities presented in Eq.~\eqref{eq:DPDF-Q0-2}, one can express the diffractive PDFs in terms of the following set of basis functions for the quark and gluon diffractive PDFs,

%--------------------------------------
\begin{eqnarray}\label{eq:DPDF-Q0-3}
\beta \, {\cal F}_q (\beta, Q_0^2; x_{\pom}) && =  {\cal W} (x_{\pom}) \nonumber \\ && {\cal N}_q \, \beta^{\alpha_q} (1 - \beta)^{\beta_q} ( 1 + \gamma_q \sqrt{\beta} + \eta_q \beta^2 )\,,  \nonumber \\
\beta \, {\cal F}_g (\beta, Q_0^2; x_{\pom}) && = {\cal W} (x_{\pom}) \nonumber \\ &&  {\cal N}_g \, \beta^{\alpha_g} (1 - \beta)^{\beta_g} ( 1 + \gamma_g \sqrt{\beta} + \eta_g \beta^2 ) \,.  \nonumber \\
\end{eqnarray}
%--------------------------------------

As already emphasized, the introduction of such flexible parameterization for the diffractive PDFs at the reference scale gives a much better description of the diffractive DIS data. 
Eq.~\eqref{eq:DPDF-Q0-3} states that the variables $x_{\pom}$, which related to the loosely scattered proton in diffractive DIS, are factorized from the variables characterizing the diffractive system $(\beta, Q^2)$. As we mentioned in Sec.~\ref{sec:factorization}, for a fixed value of $x_{\pom}$, the Q$^2$ evolution of diffractive PDFs $\beta \, {\cal F} (\beta, Q^2; x_{\pom})$ is given by the DGLAP equations.

%=====================================================================================
\section{ Diffractive DIS data sets }\label{sec:data}
%=====================================================================================

We strive to include as much of the available diffractive DIS data as possible in our data sets.
However, one need to apply some certain kinematical cuts in order to ensure that only the data sets for which the available
perturbative QCD treatment is adequate are included in the QCD fit. All the data sets used in the {\tt GKG18} diffractive PDFs analysis are also included in this analysis. 
Statistically, most significant data set that we use in our QCD analysis are the HERA measurements of the diffractive DIS ``reduced'' cross sections. 
An overview of all available and most up-to-date diffractive DIS data set is presented in Table.~\ref{tab:DDISdata}, where we indicate, for each data set: the name of experiments and the corresponding references, observables, the kinematic range of $\beta$, $x_{\pom}$ and $Q^2$, and finally the number of data points before kinematic cuts.

%
%----------------------------------
\begin{table*}[htb]
\caption{\small List of all available and most up-to-date diffractive DIS data points (before applying the kinematical cuts) used in {\tt HK19-DPDF} global analysis. For each data set we have provided the experiment and the corresponding reference, the kinematical coverage of $\beta$, $x_{\pom}$, and $Q^2$ and the number of data points. The details of each experiment explained in the text.} \label{tab:DDISdata}
\begin{tabular}{l c c c c c c c c}
		Experiment & Observable & [$\beta_{\text{min}}, \beta_{\text{max}}$] & [$x_{\pom}^{\text{min}}, x_{\pom}^{\text{max}}$]  & $Q^2\,[{\text{GeV}}^2]$  & \# of points & ${\cal N}^{\tt NLO}$   &   ${\cal N}^{\tt NNLO}$
		\tabularnewline
		\hline\hline
		H1/ZEUS Combined~\cite{Aaron:2012hua} & $\sigma_r^{D(3)}$  &  [$0.0018$, $0.562$] & [$3.0 \times 10^{-4}$, $9.0 \times 10^{-2}$] & 2.5--200 & \textbf{181} &  0.9997  &  0.9996  \\  \hline	
		H1-LRG-12~\cite{Aaron:2012ad} & $\sigma_r^{D(3)}$ & [$0.0017$, $0.80$]   & [$3.0 \times 10^{-4}$, $3.0 \times 10^{-2}$] & 3.5--1600 & \textbf{267}  &  0.9998   &  0.9997  \\
		
		H1-LRG-11 $\sqrt{s} = 225$~\cite{Aaron:2012zz} & $\sigma_r^{D(3)}$ & [$0.033$, $0.699$]   & [$5.0 \times 10^{-4}$, $3.0 \times 10^{-3}$] & 4.0--44 & \textbf{20}   &  1.0036    &  1.0049   \\
		
		H1-LRG-11 $\sqrt{s} = 252$~\cite{Aaron:2012zz} & $\sigma_r^{D(3)}$ & [$0.089$, $0.699$]   & [$5.0 \times 10^{-4}$, $3.0 \times 10^{-3}$] & 4.0--44 & \textbf{19}   &  1.0036    &  1.0049     \\
		
		H1-LRG-11 $\sqrt{s} = 319$~\cite{Aaron:2012zz} & $\sigma_r^{D(3)}$ & [$0.089$, $0.699$]   & [$5.0 \times 10^{-4}$, $3.0 \times 10^{-3}$] & 11.5--44 & \textbf{12}   &   1.0036   &  1.0049   \\	
		\hline \hline
		\multicolumn{1}{c}{\textbf{Total data}} ~~  & ~~ & ~~ & ~~ & ~~ \textbf{ 499  }  &  &   \\  \hline
\end{tabular}
\end{table*}
%
%--------------------------------
%

We now discuss the inclusion of the HERA diffractive DIS data sets into our diffractive PDFs fit. The data sets that are included in our analysis are as follow:
The H1 and ZEUS combined diffractive DIS cross section measurement~\cite{Aaron:2012hua}, the H1-LRG-12~\cite{Aaron:2012ad} data and finally the H1-LRG-11~\cite{Aaron:2012zz} data for three different center-of-mass energies of $\sqrt{s} = 225$, $252$ and $319$ GeV.

Now we are in a position to present the details of these data sets in turn. As we mentioned, in our QCD fit, we use the most recent H1 and ZEUS
combined measurement for inclusive diffractive scattering cross sections~\cite{Aaron:2012hua}. Until recently, most of the other groups in literature that have performed global diffractive PDFs analyses do not include this combined data sets. An exception is the analysis of Ref.~\cite{Goharipour:2018yov} by {\tt GKG18-DPDF}. In the present work,
and in {\tt GKG18-DPDF}, we have added the H1/ZEUS combined data in order to present a well-determined diffractive PDFs with a reliable uncertainty. This combined data set contains complete information on diffractive DIS cross sections measurement published by the H1 and ZEUS Collaborations. The kinematic range of these combined data sets is $2.5 \, {\rm GeV}^2 < Q^2 < 200 \, {\rm GeV}^2$ in photon virtuality, $3.5 \times 10^{-4} < x_{\pom} < 9.0 \times 10^{-2}$ in proton fractional momentum loss, $1.8 \times 10^{-3} < \beta < 0.816$ in scaled fractional momentum variable and finally $0.09 < |t| < 0.55$ GeV$^2$ in squared four momentum transfer at the proton vertex. This measurement used samples
of diffractive DIS data at a center-of-mass energy of $\sqrt{s} = 318 \, {\text{GeV}}$.  The combination by H1 and ZEUS Collaborations is based on the cross sections measured
with the ZEUS LPS 1~\cite{Chekanov:2004hy}, ZEUS LPS 2~\cite{Chekanov:2008fh}, H1 FPS HERA I~\cite{Aktas:2006hx} and finally the H1 FPS HERA II~\cite{Aaron:2010aa} data sets. 

Another data sets we have used in our analysis is the recent inclusive measurement of diffractive DIS at HERA by H1 Collaboration, entitled as H1-LRG-12~\cite{Aaron:2012ad}. The measurement is restricted to the phase space region $3 < Q2 < 1600$ GeV$^2$ of the photon virtuality , the square of the four momentum transfer at the proton vertex with $|t| < 1.0$ GeV$^2$ and finally the longitudinal momentum fraction of the incident proton $x_{\pom} < 0.05$ carried by the colourless exchange.
These high statistics measurements at HERA cover the data taking periods of 1999-2000 and 2004-2007. This measurement is combined with previously published results by H1 Collaboration~\cite{Aktas:2006hy} in order to provide a single data set of diffractive DIS cross sections using the large rapidity gap (LRG) selection method.  Like for the case of H1/ZEUS combined data sets, this measurement is also presented as the reduced diffractive cross section, $\sigma_r^{D(4)} (\beta, Q^2; x_{\pom}, t)$.

Finally, we have used H1 measurements of the diffractive DIS reduced cross section~\cite{Aaron:2012zz} at center-of-mass energies of $\sqrt{s} = 225$, $252$ GeV as well as the precise measurement at $\sqrt{s} = 319$ GeV. This reduced cross section measurements in this experiment, entitled H1-LRG-11, is measured in the photon virtuality range of $4.0 < Q^2 < 44.0$ GeV$^2$ and longitudinal momentum fraction of the diffractive exchange $x_{\pom}$ of   $5 \times 10^{-4} <  x_{\pom} < 3 \times 10^{-3}$.

To ensure the validity of the well-known DGLAP evolution equations one needs to impose certain cuts on the data sets we discussed in this section.
We follow the analysis by {\tt H1-2006 DPFs}~\cite{Aktas:2006hy} and continue to use the same cuts on the diffractive DIS data, i.e. $Q^2> Q^2_{min}=8.5$ GeV$^2$ in order to make sure that higher-twist corrections (HTs) are not needed in the analysis. As an aside, we should comment on the very large $x_{\pom}$ domain. One can not impose any cut at large $x_{\pom}$, although, at present, there are essentially no diffractive DIS data available probing the $x_{\pom} > 10^{-1}$ domain. In addition to the cut on $Q^2$, we include the data with $M_X > 2$ GeV in the fit.

Having discussed the details of kinematic cuts applied on the data sets in this analysis, we are now ready to discuss the corrections need to be taken into account for the H1 and ZEUS combined data. The first correction that one needs to apply on the data sets is due to the proton dissociation background. 
We should notice here that distinct methods have been used by the H1 and ZEUS experiments at HERA, and hence, the measured cross sections are not always presented with
the corrections for proton dissociation backgrounds. In this analysis, the combined H1/ZEUS diffractive DIS cross sections are corrected by a global factor of
$1.21$ to account for the such contributions.
Another correction one needs to consider comes from the fact that the all H1-LRG data sets used in our analysis are given for the range $|t| < 1$ GeV$^2$ while the combined H1/ZEUS diffractive DIS cross sections, which is based upon proton-tagged samples, are restricted to the $t$ range of $0.09 < |t| < 0.55$ GeV$^2$. Hence, one need to consider a global normalization factor to account this correction. The extrapolations for these two distinct range of $t$ have been performed by considering an exponential $t$ dependence of the inclusive
diffractive DIS cross sections, using the H1 default value of exponential slope parameter $b = 6$ 1/GeV$^2$~\cite{Aaron:2012hua,Aaron:2010aa}. Therefore, our analysis has been carried out by considering the above corrections as well as the pre-selection kinematical cuts applied on the analyzed data sets.

It is important to emphasize here that the list of diffractive DIS data sets considered in this work contains enough information to extract diffractive PDFs from a global QCD analysis. Clearly, there are other source of important processes that will provide additional information on the diffractive PDFs in the LHC era. Among these, one could consider diffractive dijet productions~\cite{Chekanov:2009ac,Aaron:2011mp,Chekanov:2007aa,Chekanov:2007aa,Chekanov:2007rh,Aaron:2010su,Andreev:2014yra,Andreev:2015cwa}, providing useful information on the diffractive gluon density and, possibly, on the quark flavor separation respectively. As we discussed in Introduction, 
the future high energy and high luminosity machines such as LHeC and FFC-eh will measure the diffractive DIS cross sections with the highest possible precision leading to the much better constrains for the diffractive PDFs~\cite{Armesto:2019gxy}. Having discussed the diffractive DIS data set as well as the kinematic cuts that we apply, we are now turn to discuss the method of $\chi^2$ minimizations and our approach to determine the diffractive PDFs uncertainties.

%=====================================================================================
\section{ The method of $\chi^2$ minimizations and diffractive PDFs uncertainties }\label{sec:minimizations}
%=====================================================================================

In the present analysis, a $\chi^2 (p)$ minimizations method to extract the fit parameters and an error calculation based on the ``Hessian approach'' to determine the uncertainties of diffractive PDFs are applied. As discussed in literature review (e.g. see~\cite{Schmidt:2018hvu,Khalek:2018mdn,Willis:2018yln}), a precise understanding of uncertainties due to PDFs in a QCD analysis is crucial to precision studies of the standard model (SM) of particle physics~\cite{Azzi:2019yne}, as well as to searches for new physics (NP) beyond the standard model (BSM)~\cite{CidVidal:2018eel}, especially for the future high-luminosity LHC and high-energy LHC.. In turn, new measurements of SM processes from present or future high energy collider experiments such as HERA, LHC, Tevatron, RHIC and LHeC can be used to constrain the uncertainties on the PDFs in a global QCD analysis. The most complete method for obtaining well constraints from the new data sets on the PDFs would be to add the up-to-date data sets into the global QCD analysis.
Currently, the two most commonly used methods to estimate the PDFs uncertainties from a QCD analysis are the Monte Carlo approach~\cite{Giele:1998gw,Giele:2001mr} and the standard Hessian method~\cite{Pumplin:2001ct, Martin:2009iq}. The details of Monte Carlo sampling techniques and use of Neural Networks (NN) as unbiased parametrization can be found in the analysis by {\tt NNPDF} Collaboration~\cite{Ball:2012cx,Ball:2016neh,Ball:2018iqk,Ball:2017nwa,Ball:2014uwa}.

The details of ``Hessian approach'' is fully addressed in Refs.~\cite{Hou:2016sho,Khanpour:2016pph,Pumplin:2001ct,Martin:2002aw,Martin:2009iq}.
In the standard ``Hessian approach??, smaller number of` ``error sets?? are considered to obtain an estimate of the PDFs errors~\cite{Pumplin:2001ct} in QCD fits. These error sets are correspond to the plus and minus eigenvectors directions in the space of PDFs fit parameters, which finally can be used to estimate the $\chi^2$ function near their global minimum. The Hessian approach relies on a quadratic approximation for the fit parameters dependence of the $\chi^2(\{\eta_i\})$ minimization function and a linear approximation  for the parameter dependence of the observables in question.  In practice, the Hessian method works quite well for the most high energy observables in which used to determine the PDFs~\cite{Pumplin:2001ct, Martin:2009iq}. A version of this method, ``Hessian profiling'', has been included in the {\tt xFitter} package~\cite{Alekhin:2014irh}. Our previous analysis on diffractive PDFs, {\tt GKG18-DPDF}~\cite{Goharipour:2018yov}, was based on this package.

Let's discuss the $\chi^2  (\{\eta_i\})$ minimization procedure first. The goodness of fit is traditionally determined by the effective global $\chi_{\tt global}^{2}(\{\eta_i\})$ minimization algorithm that measures the quality of QCD fit between theory predictions and diffractive DIS experiments. This minimization strategy allow an extraction of independent parameters $\{\eta_i\}$ which specify the diffractive PDFs at the input scale Q$_0^2 =  2$ GeV$^2$. This function is given by,

%--------------------------------
\begin{eqnarray}\label{eq:chi1}
\chi_{\tt global}^2(\{\eta_i\}) =
\sum_{n=1} ^ {N^{\tt exp}} w_n  \, \chi_n^2(\{\eta_i\}) \,,
\end{eqnarray}
%--------------------------------

where $n$ labels a particular data set and $w_n$ is a weight factor for the $n^{\tt th}$ experiment with default value of 1~\cite{Stump:2001gu,Blumlein:2006be}. $\chi_n^{2}(\{\eta_i\})$ can be written as

%--------------------------------
\begin{eqnarray}\label{eq:chi2}
\chi_n^{2}(\{\eta_i\}) = && \left( \frac{1 -{\cal N}_n }{\Delta{\cal N}_n}\right)^2 + \nonumber \\ && \sum_{j=1}^{N_n^{data}} \left(\frac{{\cal N}_n  \, {\cal O}_{j}^{\tt exp} - {\cal T}_{j}^{\tt theory} (\{\eta_i\}) }{{\cal N}_n \, \delta {\cal O}_{j}^{\tt exp}} \right)^2\,.
\end{eqnarray}
%--------------------------------

The minimization of $\chi_{\tt global}^2(\{\eta_i\})$ function presented above is done applying the CERN program library {\tt MINUIT}~\cite{James:1975dr}. 
In Eq.~\eqref{eq:chi2}, ${\cal O}$ represents the experimental measurement, and $\delta {\cal O}$ denotes the experimental uncertainty
(statistical and systematic combined in quadrature), and ${\cal T}(\{\eta_i\})$ is the theoretical value for the $i^{\tt th}$ experimental data point which depends on the input diffractive PDFs parameters $\{\eta_i\}$. In this equation, ${\cal N}_n$ is the overall normalization factors for a given data point of experiment $n$ and the ${\Delta{\cal N}_n}$ is the experimental normalization uncertainty. The (fitted) normalization factors ${\cal N}_n$ need to be extracted along with the diffractive PDFs parameters. We minimize the $\chi_{\tt global}^2 (\{\eta_i\}) $ value with the final 7 fit parameters. The obtained normalization factors ${\cal N}$ are presented in Table.~\ref{tab:DDISdata} for each experiment.  

%
%----------------------------------
\begin{table}[htbp]
\centering
{\footnotesize
\begin{tabular}{ c | c c | c}
& H1/ZEUS Combined~\cite{Aaron:2012hua}    &     &   \\  \hline 
			$x_{\pom}$    & $\chi^2$ ({\tt NLO})    & $\chi^2$ ({\tt NNLO})    & $n^{data}$  \\ \hline 
			0.0003  &             &           &   1         \\
			0.0009  & 12.070      & 12.146    &   4         \\
			0.0025  & 6.839       & 5.379     &   8           \\
			0.0085  & 18.434      & 19.340    &   15        \\
			0.016   & 19.611      & 20.401    &   17         \\
			0.025   & 22.224      & 24.137    &   18      \\
			0.035   & 9.985       & 10.019    &   19        \\
			0.05    & 30.653      & 30.787    &   20          \\
			0.075   & 13.359      & 12.787    &   19          \\
			0.09    & 3.382       & 3.492     &   7       \\
& \textbf{ 133.401  }   & \textbf{ 135.179  }    & \textbf{ 128 }   \\    \hline 
\hline
\end{tabular} 	}
\caption{The values of $\chi^2/n^{\tt data}$ for the H1/ZEUS Combined~\cite{Aaron:2012hua} data included in our QCD analysis. More detailed discussion of the description of the individual data sets, and the definitions of $\chi^2 (\{\eta_i\})$ are contained in the text. }
\label{tab:chi2-H1-ZEUS-Combined-data}
\end{table}
%
%----------------------------------
%

%
%----------------------------------
\begin{table}[htbp]
\centering
{\footnotesize
\begin{tabular}{ c | c c | c}
& H1-LRG-12~\cite{Aaron:2012ad}    &     &   \\  \hline 
				$x_{\pom}$    & $\chi^2$ ({\tt NLO})    & $\chi^2$ ({\tt NNLO})    & $n^{data}$  \\ \hline 
				0.0005  & 7.214      & 6.968    &   2         \\
				0.001  & 14.145      & 15.073    &   19         \\
				0.003  & 53.787      &  54.073   &   50           \\
				0.01  & 26.936     & 26.128   &   62        \\
				0.03   & 30.684      & 29.609    &   57         \\
& \textbf{ 132.768  }   & \textbf{ 132.499  }    & \textbf{ 188 }   \\    \hline 
\hline
\end{tabular} 	}
\caption{The values of $\chi^2/n^{\tt data}$ for the H1-LRG-12~\cite{Aaron:2012ad} included in our QCD analysis. See the caption of Table.~\ref{tab:chi2-H1-ZEUS-Combined-data} for further details. }
\label{tab:chi2-H1-LRG-12}
\end{table}
%
%----------------------------------
%
		
%
%----------------------------------
\begin{table}[htbp]
\centering
{\footnotesize
\begin{tabular}{ c | c c | c}
& H1-LRG-11 $\sqrt{s} = 225$~\cite{Aaron:2012zz}    &     &   \\  \hline 
					$x_{\pom}$    & $\chi^2$ ({\tt NLO})    & $\chi^2$ ({\tt NNLO})    & $n^{data}$  \\ \hline 
					0.03   & 16.541      & 14.493    &   13         \\
					& \textbf{ 16.541  }   & \textbf{ 14.493  }    & \textbf{ 13 }   \\    \hline 
\hline
\end{tabular} 	}
\caption{The values of $\chi^2/n^{\tt data}$ for the H1-LRG-11 $\sqrt{s} = 225$~\cite{Aaron:2012zz} included in our QCD analysis. See the caption of Table.~\ref{tab:chi2-H1-ZEUS-Combined-data} for further details. }
\label{tab:chi2-H1-LRG-11-225}
\end{table}
%
%----------------------------------
%
			
%
%----------------------------------
\begin{table}[htbp]
\centering
{\footnotesize
\begin{tabular}{ c | c c | c}
& H1-LRG-11 $\sqrt{s} = 252$~\cite{Aaron:2012zz}    &     &   \\  \hline 
						$x_{\pom}$    & $\chi^2$ ({\tt NLO})    & $\chi^2$ ({\tt NNLO})    & $n^{data}$  \\ \hline 
						0.0005  & 1.717      & 1.802    &   2         \\
						0.003   & 15.951     & 15.446    &   10         \\
     					& \textbf{ 17.669  }   & \textbf{ 17.249  }    & \textbf{ 12 }   \\    \hline 
\hline
\end{tabular} 	}
\caption{The values of $\chi^2/n^{\tt data}$ for the H1-LRG-11 $\sqrt{s} = 252$~\cite{Aaron:2012zz} included in our QCD analysis.  See the caption of Table.~\ref{tab:chi2-H1-ZEUS-Combined-data} for further details. }
\label{tab:chi2-H1-LRG-11-252}
\end{table}
%
%----------------------------------
%
			
%
%----------------------------------
\begin{table}[htbp]
\centering
{\footnotesize
\begin{tabular}{ c | c c | c}
& H1-LRG-11 $\sqrt{s} = 319$~\cite{Aaron:2012zz}    &     &   \\  \hline 
							$x_{\pom}$    & $\chi^2$ ({\tt NLO})    & $\chi^2$ ({\tt NNLO})    & $n^{data}$  \\ \hline 
							0.0005  & 0.073      & 0.229    &   2         \\
							0.003   & 8.236      & 7.793    &   10         \\
							& \textbf{ 8.309  }   & \textbf{8.022  }    & \textbf{ 12 }   \\    \hline 
\hline
\end{tabular} 	}
\caption{The values of $\chi^2/n^{\tt data}$ for the H1-LRG-11 $\sqrt{s} = 319$~\cite{Aaron:2012zz} included in our QCD analysis.  See the caption of Table.~\ref{tab:chi2-H1-ZEUS-Combined-data} for further details. }
\label{tab:chi2-H1-LRG-11-319}
\end{table}
%
%----------------------------------
%

In order to show the effects arising from the use of the different diffractive DIS data sets, in Tables.~\ref{tab:chi2-H1-ZEUS-Combined-data}, \ref{tab:chi2-H1-LRG-12}, \ref{tab:chi2-H1-LRG-11-225}, \ref{tab:chi2-H1-LRG-11-252} and \ref{tab:chi2-H1-LRG-11-319}, we present the $\chi^2$ for each bin of $x_{\pom}$ for our NLO QCD analysis. These tables illustrate the quality of our NLO and NNLO QCD fits to diffractive DIS cross sections data sets in terms of the individual $\chi^2$ values obtained for each experiment at a certain value of $x_{\pom}$. As shown in these tables all the data sets are well-fitted.  We find $\chi^{2}/{\rm d.o.f}= 312.07/346 = 0.91 \ ({\tt NNLO})$ and  $\chi^{2}/{\rm d.o.f} = 310.83/346 = 0.89 \ ({\tt NLO})$ which yield an acceptable fit to the experimental DIS data. The obtained $\chi^{2}/{\tt d.o.f}$ shows a slightly improvement in the quality of the fit at NNLO accuracy.

In the following, we briefly discuss how to apply the ``Hessian method'' to determine the uncertainties of extracted diffractive PDFs. The basic procedure of this method is provided in Refs.~\cite{Hou:2016sho,Martin:2002aw,Martin:2009iq,Pumplin:2001ct}. The diffractive PDFs, $\beta {\cal F} (\beta, Q^2; x_{\pom})$, defined at the initial scale $Q_0^2$, are parametrized by $\eta$ parameters. As we mentioned, the determination of the diffractive PDFs is obtained using a $\chi_n^{2} (\{\eta_i\})$ function given in Eq.~\eqref{eq:chi2}, which quantifies the discrepancy between the QCD theory predictions and the experimental observables of a global set of experiments, including the experimental errors.

In the well-known ``Hessian'' method one can diagonalize the covariance matrix and work in terms of the eigenvectors and eigenvalues.
As we have mentioned earlier, the appropriate parameters set can be obtained by minimizing the $\chi_n^{2} (\{\eta_i\})$ function. We entitled this as diffractive PDFs set
$s_0$. The parameters values of $s_0$, i.e. \{$\eta_1^0 \dots \eta_n^0$\}, in which extracted from QCD fit to diffractive DIS data, will be presented and discussed in details in Sec.~\ref{sec:FitResults}.

By moving away the parameters from their best fitted values, $\chi^2$ increases by the amount of $\Delta  \chi^2$
%----------------------------------
\begin{eqnarray}
\label{delta-chi} 
\Delta \chi^2_{\tt global} &=&
\chi^2_{\tt global}(\{\eta\}) -
\chi^2_0(\{\eta^0\}) \nonumber \\
&& = \sum_{i, j=1}^n (\eta_i -
\eta_i^0)  H_{ij}  (\eta_j - \eta_j^0)\,,
\end{eqnarray}
%----------------------------------
where $H_{ij}$ is the Hessian or error matrix which is given by
%----------------------------------
\begin{eqnarray}
H_{ij} = \frac{1} {2} \, 
\frac{\partial^2 \chi^2_{\tt global}}
{\partial \eta_i \,
\partial \eta_j} \Bigg|_{\tt min}\,.
\end{eqnarray}
%----------------------------------
These covariance or Hessian matrix can be obtained by running the CERN program library {\tt MINUIT}~\cite{James:1975dr}.
Having ay hand the derivative of the observable ${\cal O}$ with respect to each parameter $\{\eta\}$, one can use the following equation
%----------------------------------
\begin{eqnarray}
\label{Delta-O}
\Delta {\cal O}^{2} = 
\Delta \chi^2_{\tt global} \, \sum_{i, j = 1}^{k}
\frac{\partial {\cal O}}{\partial \eta_i}
\, C_{i j} \, \frac{\partial {\cal O}}
{\partial \eta_{j}}\,,
\end{eqnarray}
%----------------------------------
and can calculate the diffractive PDFs symmetric error bands as well as the corresponding observables such as the diffractive DIS cross sections for a desired values of confidence interval, $T = \Delta  \chi^2_{\tt global}$. We have to mentioned here that, in Eq.~\eqref{Delta-O}, $C_{ij} \equiv H_{ij}^{-1}$  is the elements of covariance or error matrix.

We should stressed that it is convenient to work in term of the eigenvalues and orthogonal eigenvectors of covariance matrix which is given by
%----------------------------------
\begin{eqnarray}
\sum_{j = 1}^n C_{ij}
\upsilon_{j k} =
\lambda_{k} \upsilon_{i k}\,.
\end{eqnarray}
%----------------------------------
The displacement of the parameter \{$\eta_i$\} from its obtained minimum values $\eta_i^0$ can be expressed in terms of the rescaled eigenvectors $e_{i  k} = \sqrt {\lambda_k}   \,  v_{i  k}$. It reads
%----------------------------------
\begin{eqnarray}
\label{zk}
\eta_{i} - \eta_{i}^{0} =
\sum_{k = 1}^n
\, e_{i k} \, z_{k} \,.
\end{eqnarray}
%----------------------------------
Considering the orthogonality of eigenvectors $\upsilon_{i k}$ and putting Eq.~\eqref{zk} in \eqref{delta-chi}, one can obtain
%----------------------------------
\begin{eqnarray}
\label{delta-chi5} 
\Delta \chi^2_{\tt global} =
\chi^{2}_{\tt global}(\{\eta\}) - 
\chi^2_0(\{\eta^0\})
= \sum_{k=1}^{n} \, z_{k}^{2}\,.
\end{eqnarray}
%----------------------------------
The relevant neighborhood of $\chi ^ 2$ is the interior of hypersphere with radius $T$.
The uncertainty in the diffractive PDFs is then set by the requirement $\Delta \chi^{2}_{\tt global}  <  T^{2}$ at some prescribed confidence level (CL). This means that
%----------------------------------
\begin{eqnarray}
\label{tt}
\sum_{k = 1}^n
z_k^2 \leq T^{2}\,.
\end{eqnarray}
%----------------------------------
Finally the neighborhood parameters
can be written as
%----------------------------------
\begin{eqnarray}
\label{t}
\eta_i(s_k^{\pm}) =
\eta_{i}^{0} \pm t \, 
\sqrt {\lambda_{k}} \,
v_{i k} \,,
\end{eqnarray}
%----------------------------------
with $s_{k}$ is the $k^{\tt th}$ set of diffractive PDFs, $t$ adapted to make the desired $T^{2} = \Delta \chi^{2}_{\tt global}$ which is the tolerance for the
required confidence interval (CL) and $t=T$ in the quadratic approximation.

Using the method discussed above, we accompany the construction of our QCD fit by reliable estimation of diffractive PDFs uncertainty.
Finally uncertainty of a given observables ${\cal O}$ in the Hessian method, which can be the diffractive PDFs or or related sort of diffractive observables,  can calculate as~\cite{Martin:2002aw,Martin:2009iq,Pumplin:2001ct}
%----------------------------------
\begin{eqnarray}\label{Delta-F1}
\Delta {\cal O} =
\frac{1}{2} \left[\sum_{k = 1}
^{n}({\cal O}(s_{k}^{+}) 
- {\cal O}(s_{k}^{-}))^{2}\right]
^{\frac{1} {2}} \,.
\end{eqnarray}
%----------------------------------
In above equation, ${\cal O}(s_{k}^{+})$ and ${\cal O}(s_{k}^{-})$ are the value of observables ${\cal O}$ extracted from the input set of parameters $\eta_{i}(s_{k}^{\pm})$ obtained from Eq.~\eqref{t}.
The $\Delta \chi^{2}_{\tt global}$ values determine the confidence region, and it is calculated so that the confidence level (CL) becomes the one-$\sigma$-error range for a given number of parameters. In this analysis, we calculate the diffractive PDFs uncertainty with $\Delta \chi^{2}_{\tt global} = 1$ and present their symmetric errors. We should mentioned here that a one-unit tolerance of $T=1$ would be the most appropriate choice if all the uncertainties
in the data points as well as the theory would be gaussian and perfectly accounted for. Hence, in order to obtain a more conservative error estimate one need to choose a reliable choice of tolerance. 

For the case of $\eta$ degrees of freedom, the $\Delta\chi^{2}_{\tt global}$ value needs to be calculated to determine the appropriate size of diffractive PDFs uncertainty. Assuming that the $\Delta\chi^{2}_{\tt global}$ follows the $\chi^{2}$ distribution with $\eta$ degrees of freedom, we have the confidence level P as~\cite{Martin:2009iq,Martin:2002aw}
%
%----------------------------------
%----------------------------------
\begin{eqnarray}\label{eq:deltachi2}
{\text{P}} =
\int_{0}^{\Delta\chi^{2}_{\tt global}}
\frac{1}{2 \Gamma(\eta/2)
}(\frac{\xi}{2})^{(\eta/2)-1}
e^{-\frac{\xi}{2}} \, d\xi \,,
\end{eqnarray}
%----------------------------------
%
where $\Gamma$ is the Gamma function. For the case of the one-free-parameter fit, we obviously have $\Delta\chi^{2}_{\tt global}=1$. 
Sine the diffractive PDFs in common QCD fits are considered with several free fit parameters, $\eta>1$, the value of $\Delta\chi^{2}_{\tt global}$ should be calculated form Eq.~\eqref{eq:deltachi2}. Considering the parameter number in our diffractive PDFs analysis, Eq.~\eqref{eq:deltachi2} leads to the tolerance criterion of $\Delta \chi^{2}_{\tt global} = 5.0$

%------------------------------------------------
\begin{figure*}[htb]
	\vspace{0.20cm}
	\includegraphics[clip,width=0.70\textwidth]{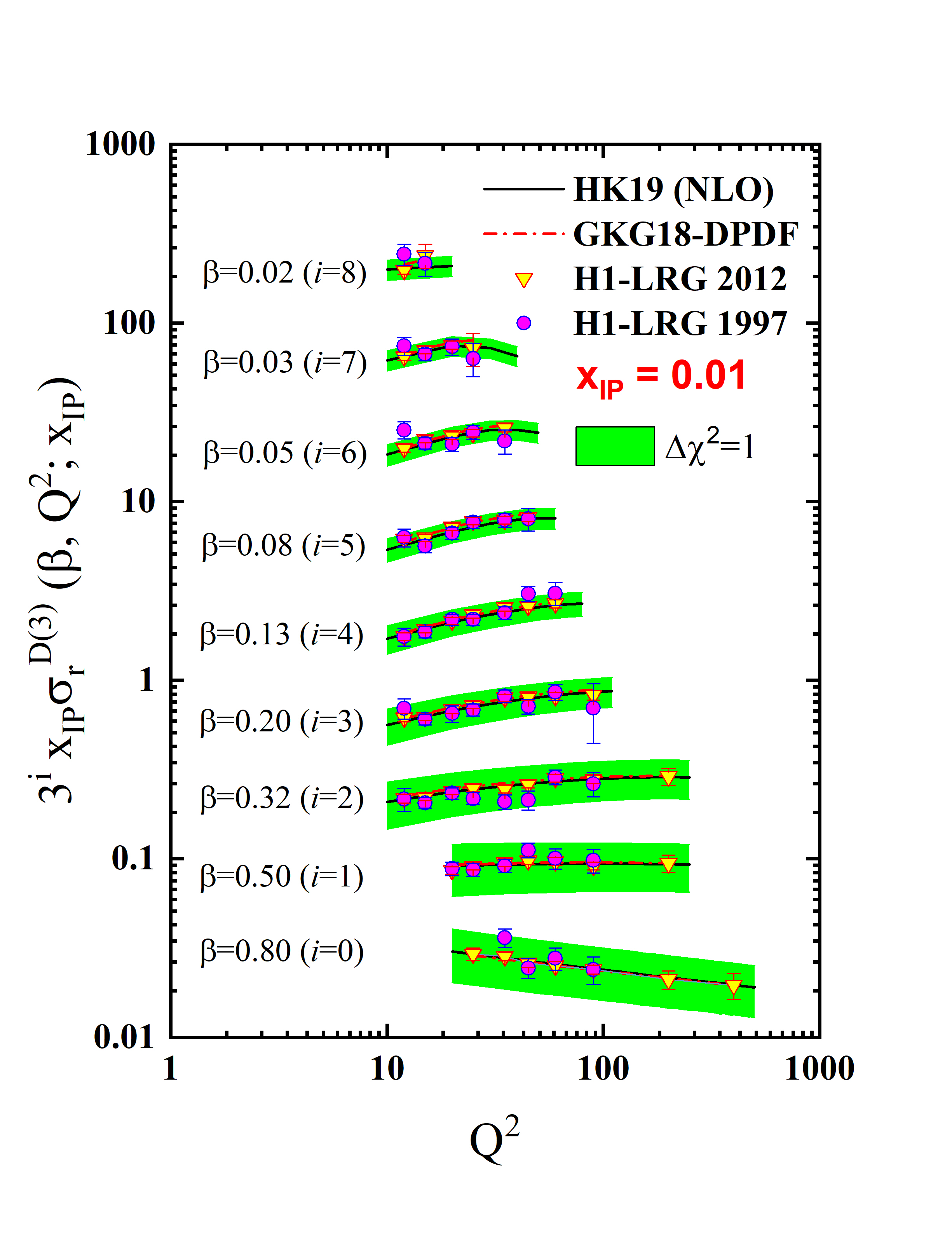}
	%\vspace{-6cm}
	\begin{center}
		\caption{{\small The NLO theory predictions for the diffractive reduced cross sections $x_{\pom} \sigma_r^{D(3)} (\beta, Q^2; x_{\pom})$ as a function of Q$^2$ for some selected values of $\beta$ and for $x_{\pom} = 0.01$. To facilitate the graphical presentation, we have plotted $3^i \times x_{\pom} \sigma_r^{D(3)}$ with $i$ indicated in parentheses in the figure. The NLO theory predictions are compared with the H1-LRG-2012~\cite{Aaron:2012ad} and H1-LRG-1997~\cite{Aktas:2006hy} measurements. The NLO theory prediction based on the {\tt GKG18-DPDF}~\cite{Goharipour:2018yov} also have been shown for comparison.  } \label{fig:xp01}}
	\end{center}
\end{figure*}
%------------------------------------------------
%------------------------------------------------
\begin{figure*}[htb]
	\vspace{0.20cm}
	\includegraphics[clip,width=0.70\textwidth]{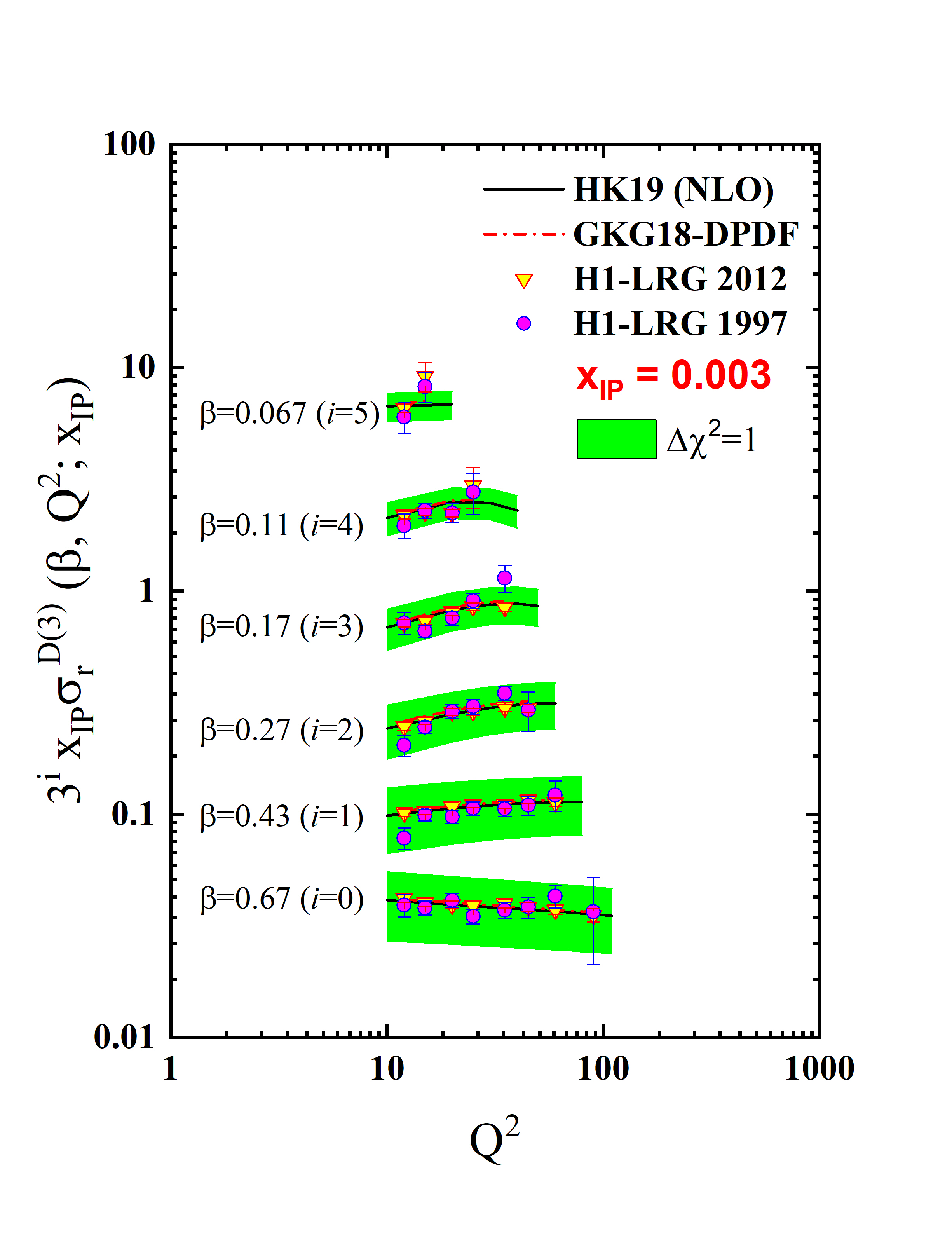}
	%\vspace{-6cm}
	\begin{center}
		\caption{{\small The NLO theory predictions for the diffractive reduced cross sections $x_{\pom} \sigma_r^{D(3)} (\beta, Q^2; x_{\pom})$ for $x_{\pom} = 0.003$. See the caption of Fig.~\ref{fig:xp01} for further details. } \label{fig:xp003}}
	\end{center}
\end{figure*}
%------------------------------------------------

%------------------------------------------------
\begin{figure*}[htb]
	\vspace{0.20cm}
	\includegraphics[clip,width=0.70\textwidth]{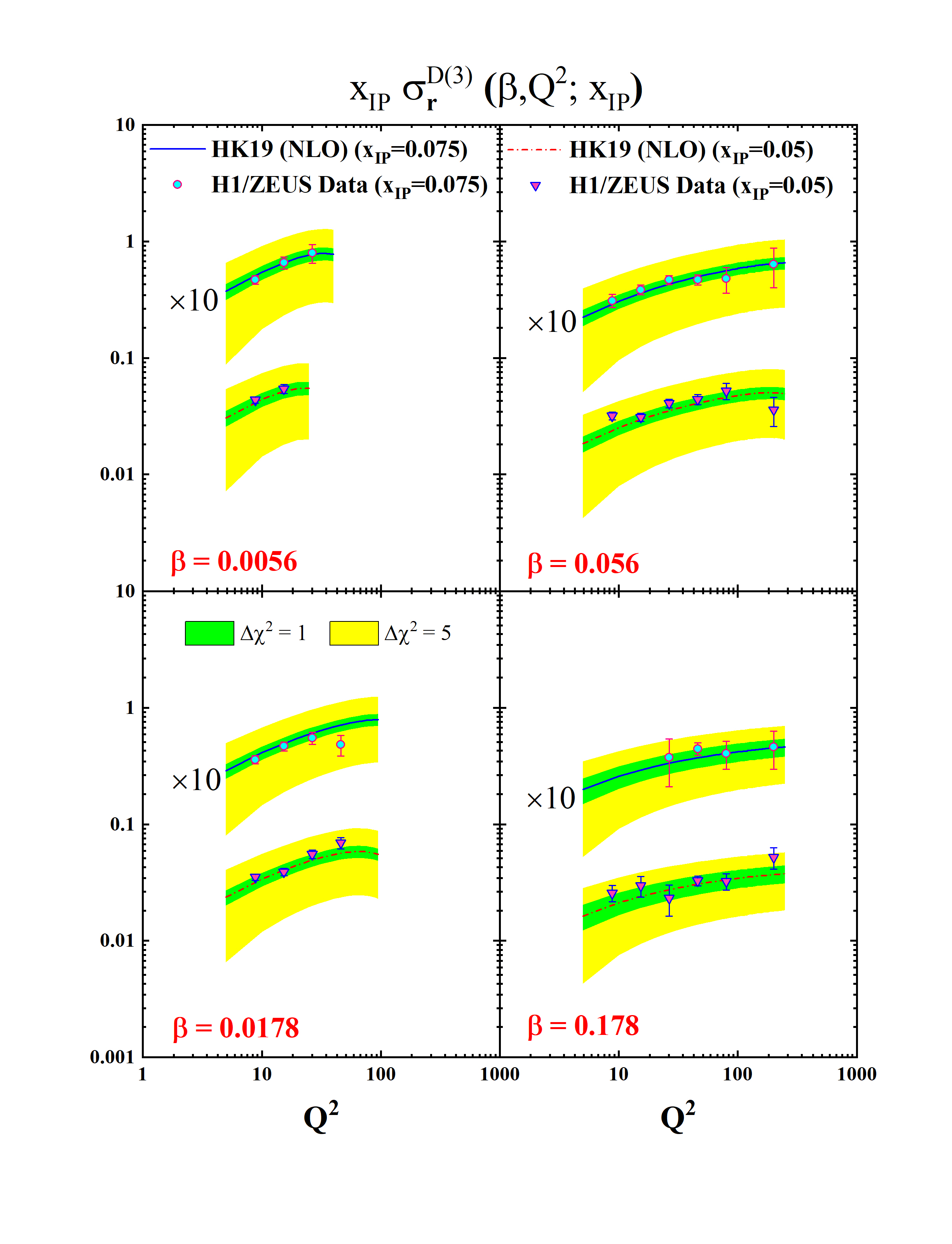}
	%\vspace{-6cm}
	\begin{center}
		\caption{{\small The NLO theory predictions for the diffractive reduced cross sections $x_{\pom} \sigma_r^{D(3)} (\beta, Q^2; x_{\pom})$ as a function of Q$^2$ for some selected values of $\beta$ and for two representative bins of $x_{\pom} = 0.05$ and 0.075. The H1/ZEUS combined~\cite{Aaron:2012hua} diffractive DIS measurements also has been shown for comparison. The error bands correspond to the fit uncertainties for the choice of tolerance $\Delta \chi^{2} = 1$ and $\Delta \chi^{2} = 5$. } \label{fig:Combined-Model-NLO}}
	\end{center}
\end{figure*}
%------------------------------------------------

%------------------------------------------------
\begin{figure*}[htb]
	\vspace{0.20cm}
	\includegraphics[clip,width=0.90\textwidth]{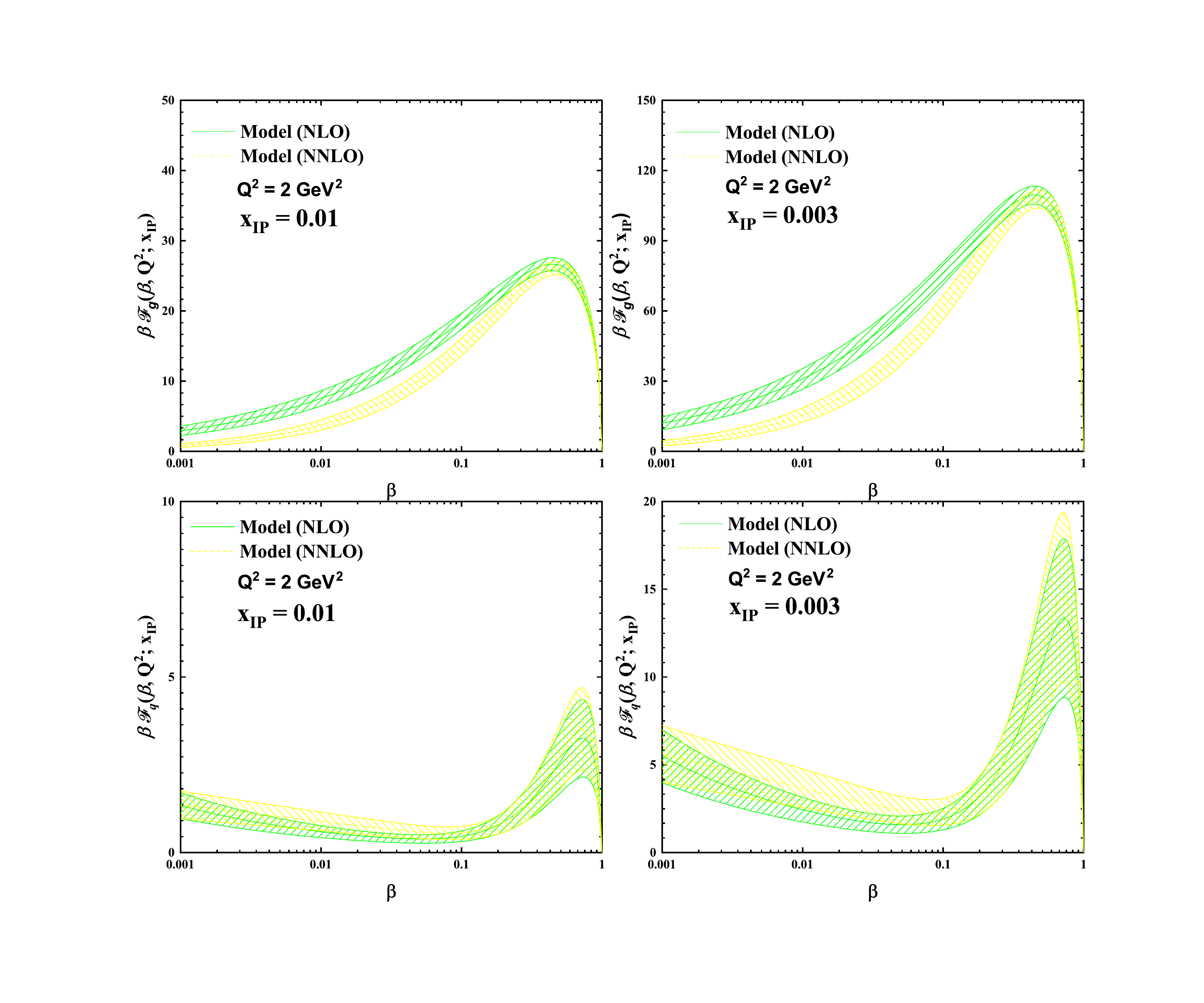}
	%\vspace{-6cm}
	\begin{center}
		\caption{{\small  Comparison the NLO and NNLO  quark  $\beta \, {\cal F}_q (\beta, Q^2; x_{\pom})$  and gluon $\beta \, {\cal F}_g (\beta, Q^2; x_{\pom})$ diffractive PDFs
				at the input scale Q$_0^2$ = 2 GeV$^2$ for two different $x_{\pom}$ bin of 0.01 and 0.003. The uncertainty bands of diffractive PDFs presented for the choice of tolerance $T=\Delta\chi^{2}_{\tt global}=1$ for the 68\% (one-sigma) confidence level (CL). } \label{fig:Q0=2GeV2-NLO-NNLO}}
	\end{center}
\end{figure*}
%------------------------------------------------

%------------------------------------------------
\begin{figure}[htb]
	\vspace{0.20cm}
	\includegraphics[clip,width=0.5\textwidth]{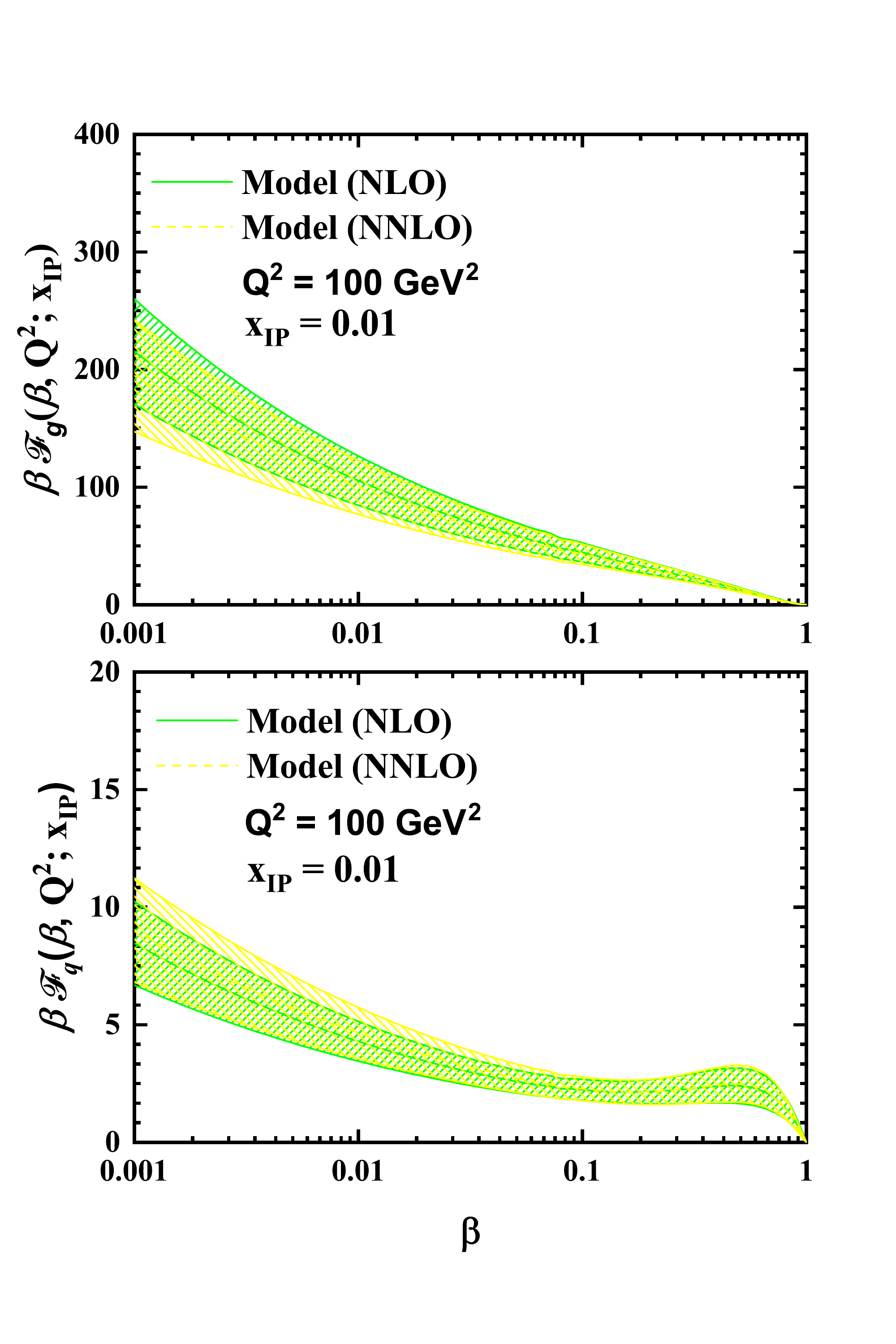}
	%\vspace{-6cm}
	\begin{center}
		\caption{{\small  Comparison the NLO and NNLO quark $\beta \, {\cal F}_q (\beta, Q^2; x_{\pom})$ and gluon $\beta \, {\cal F}_g (\beta, Q^2; x_{\pom})$ diffractive PDFs at Q$^2$ = 100 GeV$^2$ for $x_{\pom} = 0.01$. The uncertainty bands of diffractive PDFs presented for the choice of tolerance $T=\Delta\chi^{2}_{\tt global}=1$. } \label{fig:NLO-and-NNLO-Q100-xP-0-01}}
	\end{center}
\end{figure}
%------------------------------------------------

%------------------------------------------------
\begin{figure*}[htb]
	\vspace{0.20cm}
	\includegraphics[clip,width=0.90\textwidth]{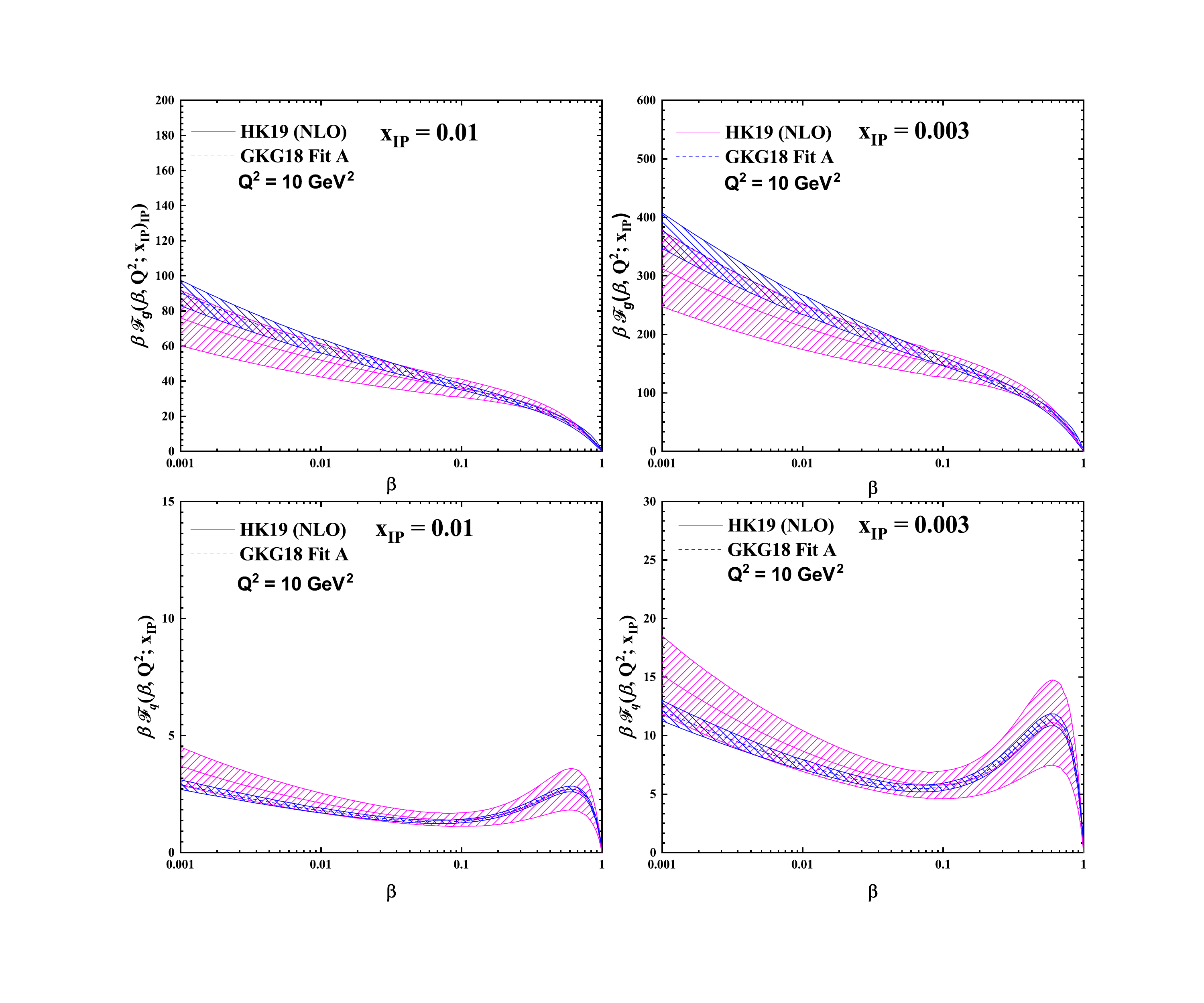}
	%\vspace{-6cm}
	\begin{center}
		\caption{{\small  Comparison between our NLO quark and gluon diffractive PDFs and the results by {\tt GKG18-DPDF}~\cite{Goharipour:2018yov}. We show results for the diffractive gluon and quark PDFs at Q$^2$ = 10 GeV$^2$ for two selected bin of $x_{\pom}$ = 0.01 and 0.003. The uncertainty bands for {\tt HK19-DPDF} results as well as for the {\tt GKG18-DPDF} analysis are correspond to the choice of tolerance $T=\Delta\chi^{2}_{\tt global}=1$.   } \label{fig:Q-10GeV-xP-0-01-0-003-Model-GKG}}
	\end{center}
\end{figure*}
%------------------------------------------------

%------------------------------------------------
\begin{figure}[htb]
	\vspace{0.20cm}
	\includegraphics[clip,width=0.5\textwidth]{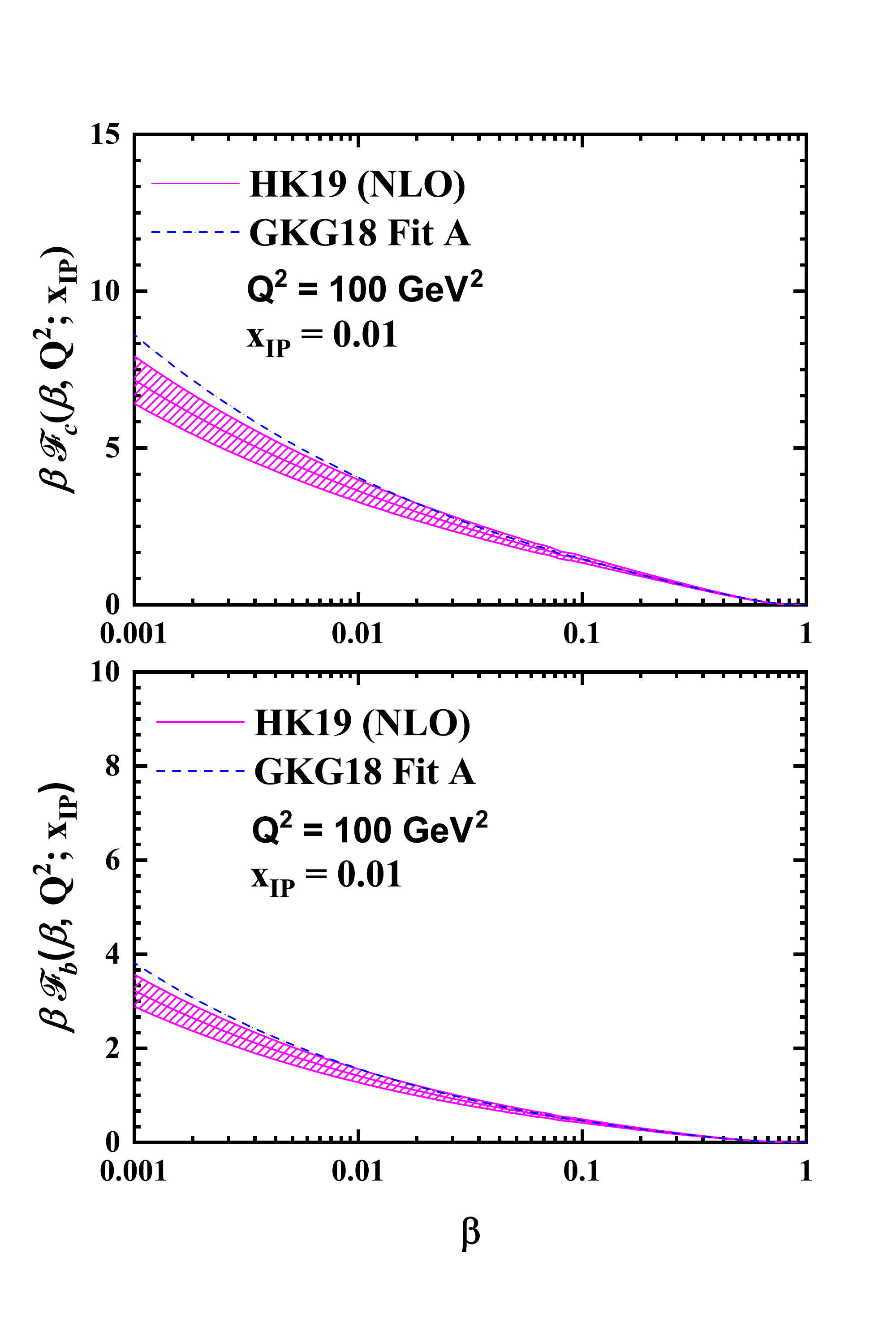}
	%\vspace{-6cm}
	\begin{center}
		\caption{{\small  Charm $\beta {\cal F}_c (\beta, Q^2; x_{\pom})$ and bottom $\beta {\cal F}_b (\beta, Q^2; x_{\pom})$ quark diffractive PDFs obtained from our NLO QCD fits at Q$^2$ = 100 GeV$^2$ and for $x_{\pom}$ = 0.01. The results from {\tt GKG18-DPDF}~\cite{Goharipour:2018yov} analysis also presented for comparison. The uncertainty bands are correspond to the choice of tolerance $T=\Delta\chi^{2}_{\tt global}=1$. } \label{fig:Charm-and-Bottom-Model-GKG}}
	\end{center}
\end{figure}
%------------------------------------------------

In the next section, we will present full details of our fit quality and the extracted diffractive PDFs.

%=====================================================================================
\section{ Results and discussions }\label{sec:FitResults}
%=====================================================================================

In this section, we discuss the main results of this work, namely the NLO and NNLO diffractive PDFs sets extracted form QCD analyses of diffractive DIS data sets. 
We first discuss the numerical results of present work including the values of fitted parameters. Secondly, we discuss the quality of the QCD fits and compare  our NLO and NNLO predictions to the fitted data sets. Then we show the resulting diffractive PDFs and their uncertainties. We mainly focus on the perturbative convergence upon applying the fracture function framework as well as inclusion of higher-order QCD corrections. Then we compare the extracted diffractive PDFs with the most recent results in literature, especially the results from {\tt GKG18-DPDF}~\cite{Goharipour:2018yov} analysis. 

In Table.~\ref{tab-dpdf}, we present the fit parameters obtained in this work for our NLO and NNLO QCD analyses at the initial scale of $Q_{0}^{2} = 2 \, {\text{GeV}}^2$. Values marked with (*) are fixed in the fit since the analyzed data sets do not constrain these parameters well enough. These fitted parameters include \{${\cal N}$, $\alpha$, $\beta$, $\gamma$, $\eta$\} for the diffractive quark and gluon densities. The parameters for the weight factor ${\cal W} (x_{\pom})$ also presented as well.
As one can see from this table, the values of the physical parameters used in the computation of diffractive DIS cross section and in the evolution of diffractive PDFs are the same as those used in the {\tt NNPDF} global analysis of unpolarized PDFs and FFs~\cite{Bertone:2017tyb,Bertone:2018ecm,Ball:2017nwa}. Specifically, we use $\alpha_s(M_Z^2) = 0.1185$ as reference value for the QCD couplings, and $m_c = 1.51$ and $m_b = 4.92$ GeV for the charm- and bottom-quark masses.

%
%--------------------------------
%
\begin{table}[ht]
\begin{center}
\caption{\small Best fit parameters obtained in {\tt HK19-DPDF} NLO and NNLO QCD fits at the initial scale of $Q_{0} ^ {2} = 2 \, {\text{GeV}}^2$ and their experimental uncertainties. Values marked with (*) are fixed in the fit since the analyzed diffractive DIS data sets do not constrain these parameters well enough. }
\begin{tabular}{ c | c | c }
\hline \hline
			Parameters	  & {\tt NLO}             & {\tt NNLO}           \\  \hline \hline
			${\cal N}_q$  & $0.0005 \pm 0.0002$   & $0.0023 \pm 0.0008$  \\ 
			$\alpha_q$    & $-0.3334 \pm 0.0388$  & $-0.1298 \pm 0.030$  \\ 
			$\beta_q$     & $0.6157 \pm 0.03264$  & $0.7308 \pm 0.032$   \\ 
			$\gamma_q$    & $-0.663^*$            & $-1.647^*$           \\ 
			$\eta_q$      & $89.327^*$            & $28.959^*$           \\     \hline  
			${\cal N}_g$  & $0.2075 \pm 0.01027$  & $0.3265 \pm 0.018$   \\
			$\alpha_g$    & $0.4144 \pm 0.03791$  & $0.662 \pm 0.0523$   \\ 
			$\beta_g$     & $0.50^*$              & $0.50^*$              \\ 
			$\gamma_g$    & $-0.0366^*$           & $-0.4200^*$       \\ 
			$\eta_g$      & $0.0^*$               & $0.0^*$               \\   \hline  
			$w_1$         & $-1.1912 \pm 0.00611$ & $-1.1969 \pm 0.0063$   \\
			$w_2$         & $0.0^*$               & $0.0^*$               \\
			$w_3$         & $86.156^*$            & $82.532^*$        \\ 
			$w_4$         & $1.7735 \pm 0.02647$  & $1.7441 \pm 0.02595$        \\ 			 \hline  \hline
			$\alpha_s(M_Z^2)$   & $0.1185^*$~\cite{Bertone:2017tyb,Bertone:2018ecm,Ball:2017nwa} & $0.1185^*$~\cite{Bertone:2017tyb,Bertone:2018ecm,Ball:2017nwa}   \\
			$m_c$               & $1.51^*$~\cite{Bertone:2017tyb,Bertone:2018ecm,Ball:2017nwa}   & $1.51^*$~\cite{Bertone:2017tyb,Bertone:2018ecm,Ball:2017nwa}     \\
			$m_b$               & $4.92^*$~\cite{Bertone:2017tyb,Bertone:2018ecm,Ball:2017nwa}   & $4.92^*$~\cite{Bertone:2017tyb,Bertone:2018ecm,Ball:2017nwa}     \\ 	\hline \hline
			$\chi^2/{\rm d.o.f}$  & $312.07/346 = 0.91$   & $310.83/346 = 0.89$  \\  	\hline \hline
\end{tabular}
\label{tab-dpdf}
\end{center}
\end{table}
%
%--------------------------------
%

In the present fit of diffractive PDFs, the analyzed data sets is much more limited than in a typical global proton PDFs fit. As we discussed earlier, diffractive DIS data allow for
the determination of only two independent combinations of densities, namely the singlet and the gluon. In addition, these distributions at some region of $\beta$ sill remain unconstrained. This effect is more enhanced for the case of diffractive gluon density, due to the reduced sensitivity of the diffractive DIS data included in our QCD fit to the gluon distribution. Therefore one would expect that the diffractive gluon PDFs is determined with larger uncertainties than the diffractive quark PDFs. This issue deserves a separate comments which we will discuss later. In addition, diffractive DIS is blind to the separation between quark and antiquark distributions. These points confirm that the input functional form and it's sensitivity to the parameter space of \{$\eta_i$\} and \{$w_i$\} need to be clearly investigated. As we mentioned, the currently
available diffractive DIS data sets do not fully constrain the entire $\beta$ and $x_{\pom}$ dependence of quark and gluon diffractive PDFs presented in Eq.~\eqref{eq:DPDF-Q0-3}. Consequently, we are forced to make some restrictions on the parameter space of \{$\eta_i$\} and \{$w_i$\}.
For the diffractive gluon density, we set $\eta_g$ to 0 and $\beta_g$ to 0.5. These only marginally limit the freedom in the input functional form for the gluon density. We fixed the $\gamma_q$, $\eta_q$ and $\gamma_g$ to their best fit values. The lack of diffractive DIS data at high-$x_{\pom}$, mean that the flux factor is not really well determined in this region. Hence, for the flux factor, we set the $w_2$ to 0 and fixed other variable \{$w_3$\} to their best fit values. In total, these leave us with 7 free parameters in our QCD fit (three for quarks, two for the gluon density and two for the flux factor), which we include later on our diffractive PDFs uncertainty estimations. 

In the following, we discuss the overall fit quality. Figs.~\ref{fig:xp01} and \ref{fig:xp003} show the NLO theory predictions based on the extracted diffractive PDFs for the diffractive reduced cross sections $x_{\pom} \sigma_r^{D(3)} (\beta, Q^2; x_{\pom})$ for two representative bins of $x_{\pom} = 0.01$ and $0.003$. The uncertainty bands correspond to the choice of tolerance $T=\Delta\chi^{2}_{\tt global}=1$ also have been shown as well. These error bands represent the
fit uncertainties derived only from the experimental input. In order to judge the fit quality, our NLO theory predictions are compared with the H1-LRG-2012~\cite{Aaron:2012ad} measurements. It would be also interesting to examine the fit quality in comparison to the data sets that we did not include in our fit such as old diffractive DIS data measurements at HERA. Hence, in Figs.~\ref{fig:xp01} and \ref{fig:xp003}, we also compare our NLO theory predictions with the H1-LRG-1997~\cite{Aktas:2006hy} measurements. As one can see, prediction based on our diffractive PDFs sets are in reasonably good agreements with the HERA data. The NLO theory prediction based on the {\tt GKG18-DPDF}~\cite{Goharipour:2018yov} also have been shown for comparison. From the theory predictions presented in these figures one can conclude that the scale dependence induced by the evolution equations of Eq.~\eqref{eq:DGLAP} is perfectly consistent with the diffractive DIS data. The results clearly indicate that one can use the fracture functions approach to describe diffractive DIS in perturbative QCD at the kinematic region covered by the $ep$ collider HERA as well as other hadron colliders.

% Fig

The previous figures show the quality of the description of the H1-LRG-2012 data with our NLO theory predictions. It would be interesting to observe that the fit quality to the inclusive H1 and ZEUS combined measurement for the inclusive diffractive DIS cross sections. In the following we discuss the overall fit quality of this data sets. For completeness, in Fig.~\ref{fig:Combined-Model-NLO}, the NLO theory predictions for the diffractive reduced cross sections $x_{\pom} \sigma_r^{D(3)} (\beta, Q^2; x_{\pom})$ have been presented as a function of Q$^2$ for some selected values of $\beta$ and for two representative bins
of $x_{\pom} = 0.05$ and 0.075. The H1/ZEUS combined~\cite{Aaron:2012hua} diffractive DIS measurements also has been shown for comparison. The uncertainty bands for {\tt HK19-DPDF} analysis as well as for the {\tt GKG18-DPDF} are correspond to the choice of tolerance $T=\Delta\chi^{2}_{\tt global}=1$ and $T=\Delta\chi^{2}_{\tt global}=5$.  These results show that the quality of the description between our NLO theory predictions and all the H1/ZEUS combined data points analyzed in this study is quite acceptable.

% Fig

In the rest of this section, we present the resulting diffractive PDFs at NLO and NNLO accuracy. We also compare the {\tt HK19-DPDF} results with the most recent analysis of {\tt GKG18-DPDF}~\cite{Goharipour:2018yov}. We mainly discuss the main difference of the results as well as the effect arising from including the higher order QCD corrections. 
We now compare the diffractive PDFs $\beta \, {\cal F} (\beta, Q^2; x_{\pom})$ obtained from this analysis with other results in literature, and discuss how their central values and uncertainties vary. As we presented in Sec.~\ref{sec:minimizations}, this analysis is based on a standard ``parameter-fitting'' criterion, and we plan to present the uncertainty bands of diffractive PDFs considering the choice of tolerance $T=\Delta\chi^{2}_{\tt global}=1$ for the 68\% (one-sigma) confidence level (CL) uncertainty.

In the following, we now turn to discuss the obtained diffractive PDFs and their uncertainties. In Fig.~\ref{fig:Q0=2GeV2-NLO-NNLO} detailed comparisons of the NLO and NNLO diffractive quark  $\beta \, {\cal F}_q (\beta, Q^2; x_{\pom})$  and gluon $\beta \, {\cal F}_g (\beta, Q^2; x_{\pom})$ PDFs have been presented at the input scale Q$_0^2$ = 2 GeV$^2$ for two different $x_{\pom}$ bin of 0.01 and 0.003. The uncertainty bands of diffractive PDFs presented for the choice of tolerance $T=\Delta\chi^2_{\tt global}=1$ for the 68\% (one-sigma) confidence level (CL). In order to study the perturbative convergence of the diffractive PDFs upon inclusion of higher order QCD corrections, in Fig.~\ref{fig:Q0=2GeV2-NLO-NNLO} we also compare our NLO and NNLO determinations among each other. Comparing our results, it can be seen that the NLO and NNLO quark diffractive PDFs $\beta \, {\cal F}_q$ are similar in size for all range of $\beta$. However a small difference can be seen in range of $\beta < 0.1$ for the $x_{\pom}$ = 0.003. The single most striking observation to emerge from the NLO and NNLO comparisons is for the case of the diffractive gluon PDFs $\beta \, {\cal F}_g$. One can see that the diffractive gluon PDFs is affected by including the higher-order QCD corrections. Concerning the shapes of the diffractive gluon PDFs, a number of interesting differences between the NLO and NNLO results can be seen from the comparisons in Fig.~\ref{fig:Q0=2GeV2-NLO-NNLO}. Significant differences in shape are observed for the small values of $\beta$, especially for the range of $\beta < 0.2$. In this range, the diffractive gluon PDFs at NLO accuracy have a larger magnitude than the NNLO results.

As one can see the NLO and NNLO uncertainties are similar in size showing that the inclusion of higher-order QCD corrections do not improve the uncertainty.  
As we presented in Tables.~\ref{tab:chi2-H1-ZEUS-Combined-data} to \ref{tab:chi2-H1-LRG-11-319}, concerning the fit quality of the total diffractive DIS data set, the most noticeable feature is the improvement upon inclusion of higher-order corrections. However, the improvement of the total $\chi^2/{\rm d.o.f}$ is not significant when going from NLO to NNLO. This demonstrates that the inclusion of the NNLO QCD corrections could slightly improves the fit quality as well as the description of the data.

In Fig.~\ref{fig:NLO-and-NNLO-Q100-xP-0-01}, we present the quark $\beta \, {\cal F}_q (\beta, Q^2; x_{\pom})$ and gluon $\beta \, {\cal F}_g (\beta, Q^2; x_{\pom})$ diffractive PDFs
at the scale of Q$^2$ = 100 GeV$^2$  and for the $x_{\pom} = 0.01$. Comparing these two results, it can be seen that the differences between the NLO and NNLO sets are very small, both for central values and uncertainties. 

% Fig

We are in a position to compare our best-fit NLO diffractive PDFs to their counterparts in the {\tt GKG18-DPDF} analysis~\cite{Goharipour:2018yov} which applied the general method used to extract diffractive PDFs from available data by considering a number of assumptions motivated by the Regge phenomenology. In Fig.~\ref{fig:Q-10GeV-xP-0-01-0-003-Model-GKG}, we present the results for the diffractive gluon and quark PDFs at Q$^2$ = 10 GeV$^2$ for two selected bin of $x_{\pom}$ = 0.01 and 0.003. We see that the overall effects of the new methodology are comparable to those reported by~\cite{Goharipour:2018yov}, but that they act in different kinematical regions of $\beta$ and $x_{\pom}$ and for different diffractive PDFs. From Fig.~\ref{fig:Q-10GeV-xP-0-01-0-003-Model-GKG}, it is clear that central values move very little while diffractive PDFs uncertainties are slightly increased. For instance, for the light quarks and the gluon diffractive PDFs the impact of the new methodology mainly effect the small regions of $\beta$; $\beta < 0.1$, where it produces an enhancement for the gluon density and a reduction for the quark density. We should emphasize here that, the QCD analysis of diffractive PDFs motivated by the Regge phenomenology uses the GRV parametrization derived from a fit to pion structure function data~\cite{Gluck:1991ng} for the Reggeon parton density, and hence, an additional source of uncertainty need to be taken into account in presenting the diffractive PDFs in the kinematic space of $z$, Q$^2$ and $x_{\pom}$. Overall, there is satisfactory agreement between the two methodology of diffractive PDFs determinations. Hence, this study strengthens the idea of using the fracture function approach to determine diffractive PDFs from a QCD analysis of diffractive DIS data sets.

% Fig

Let us conclude this section with a presentation of heavy quark diffractive PDFs obtained in this study. As we discussed in section~\ref{sec:heavy-flavour}, for the calculation of heavy-quark structure functions is performed in the {\tt FONLL} GM-VFNS~\cite{Forte:2010ta}. In Fig.~\ref{fig:Charm-and-Bottom-Model-GKG}, the charm $\beta  c (\beta, Q^2; x_{\pom})$ and bottom $\beta b (\beta, Q^2; x_{\pom})$ quark diffractive PDFs obtained from our NLO QCD fits have been shown at Q$^2$ = 100 GeV$^2$ and for $x_{\pom}$ = 0.01. The error bands shown in these figures ar correspond to the fit uncertainties derived only from the experimental input. The results from {\tt GKG18-DPDF} analysis~\cite{Goharipour:2018yov} also presented for comparison. As one can see from these results, the overall agreements are well and only small differences between our NLO results and {\tt GKG18-DPDF} can be found for all heavy quark diffractive PDFs at lower values of $\beta$; $\beta < 0.02$.

% Fig

Through our results presented in section, we have shown that our QCD analyses show good agreements with the results obtained by {\tt GKG18-DPDF} parameterization. In addition, we found that our theory predictions based on the extracted diffractive PDFs are in satisfactory agreements with the H1 and ZEUS diffractive DIS data sets over a wide range of DIS kinematics. The scale dependence of these data sets is found to be in satisfactory agreement with the one predicted for the fracture function, driven by the DGLAP evolution equations. As a short summary, the results of this research support the idea of extracting the diffractive PDFs using the fracture functions approach. 
The presented QCD-based predictions of diffractive DIS processes also provide an important step towards an improved understanding of such processes and also represent a precise test of the employed theoretical concepts to extract PDFs from a QCD analysis of diffractive DIS observables. In particular, it is observed in this analysis, that for the given kinematical range of the HERA hard diffraction events, higher-order QCD corrections are of crucial importance.

%
%%%%%%%%%%%%%%%%%%%%%%%%%%%%%%%%%%%%%%%%%%%%%%%%%%%%%%%%%%%%%%%%%%%%%%
\section{Summary and conclusions} \label{sec:Discussion}
%%%%%%%%%%%%%%%%%%%%%%%%%%%%%%%%%%%%%%%%%%%%%%%%%%%%%%%%%%%%%%%%%%%%%%
%

In summary, in this work, we have presented a set of diffractive PDFs at next-to-leading order (NLO) and next-to-next-leading order (NNLO) accuracy obtained from the most up-to-date diffractive DIS data including the most recent combined data sets from H1 and ZEUS Collaborations. The new combined diffractive DIS data from run II at HERA allow for a very accurate determination of the quark and gluon distributions in a wide range of scaled fractional
momentum $\beta$ and longitudinal momentum fractions $x_{\pom}$. We supplement our best-fit diffractive PDFs parameterizations with the reliable uncertainties obtained according to the ``Hessian approach'' which allows the experimental uncertainties to propagate to an arbitrary observable such as diffractive DIS cross sections. The theory predictions for the hard-scattering diffractive DIS processes in this analysis maintain the NLO accuracy, and for the first time the NNLO accuracy in QCD and employ the {\tt FONLL} GM-VNFS, which has been shown to provide an excellent description of the existing diffractive DIS data.

In addition to the points mentioned, the theory framework applied in this analysis features a number of new improvements. We work in the framework of fracture functions, as a new method to extract diffractive PDFs inspired by the fully factorization theorem for diffractive DIS processes, which allows a good description of diffractive DIS cross sections. We have demonstrated that diffractive DIS is consistent within this picture and one can introduced diffractive PDFs accordingly.
We have shown that a simple parameterization form for the diffractive PDFs along with the fully factorized approach for the cross section provide very accurate descriptions of the diffractive DIS data sets measured by H1 and ZEUS collaborations at HERA. The diffractive PDFs extracted for a QCD analysis in the fracture functions are also in satisfactory agreements by other analysis in literature. Finally, our results verify that the scale dependence of the data agrees well with the one predicted by the fracture function formalism. Hence, we can conclude that in the diffractive DIS kinematics analyzed in this study, the fracture functions framework can provide a good understanding of the physical picture of this sort of high energy processes. 

Our analysis can be extended in various different directions. First, our analysis can be extended to the diffractive dijet events at HERA~\cite{Andreev:2015cwa,Andreev:2014yra}. For the future, our main goal is to assess the impact of diffractive dijet productions data on the diffractive PDFs and their uncertainties. More detailed discussions on this new extraction of diffractive PDFs at NLO accuracy will be presented in our next study. Second, in terms of future work, it would be interesting to repeat the analysis described here and present a combined QCD analysis of recent data sets measured by the H1 and ZEUS collaborations at HERA on the diffractive DIS and leading-nucleon productions~\cite{Aaron:2010ab,Chekanov:2002pf,Chekanov:2008tn}. The LO, NLO and NNLO diffractive PDFs sets $\beta \, {\cal F} (\beta, Q^2; x_{\pom})$ presented in this work are available in the {\tt LHAPDF} format~\cite{Buckley:2014ana} from the author upon request.

%
%%%%%%%%%%%%%%%%%%%%%%%%%%%%%%%%%%%%%%%%%%%%%%%%%%%%%%%%%%%%%%%%%%%%%%%
\begin{acknowledgments}
%%%%%%%%%%%%%%%%%%%%%%%%%%%%%%%%%%%%%%%%%%%%%%%%%%%%%%%%%%%%%%%%%%%%%%%
%

The author is especially grateful to Muhammad Goharipour, Maryam Soleymaninia and S. Atashbar Tehrani for carefully reading the manuscript and fruitful discussions. Author thanks University of Science and Technology of Mazandaran, and School of Particles and Accelerators, Institute for Research in Fundamental Sciences (IPM) for financial support of this project.

\end{acknowledgments}
%

%%%%%%%%%%%%%%%%%%%%%%%%%%%%%%%%

%
% BibTeX users please use
% \bibliographystyle{}
% \bibliography{}
%
% Non-BibTeX users please use

 \clearpage

%
%%%%%%%%%%%%%%%%%%%%%%%%%%%%%%%%

\end{document}